\documentclass[aps,prd,twocolumn,nofootinbib]{revtex4}
\usepackage{graphicx}
\usepackage{hyperref}
\usepackage{slashed}
\usepackage{color}
\usepackage{subfig}

\begin{document}

\title {Excited heavy quarkonium production in Higgs boson decays}

\author{Qi-Li Liao$^{1}$}
\email{xiaosueer@163.com}
\author{Jun Jiang$^{2}$}
\email{jiangjun87@sdu.edu.cn}

\address{$^{1}$College of Mobile Telecommunications, Chongqing University of Posts and Telecommunications, Chongqing 401520, China \\$^{2}$ School of Physics, Shandong University, Jinan 250100, Shandong, China}

\date{\today}

\begin{abstract}
The rare decay channels of Higgs boson to heavy quarkonium offer vital opportunities to explore the coupling of Higgs to heavy quarks. We study the semi-exclusive decay channels of Higgs boson to heavy quarkonia, i.e., $H^0\to |(Q\bar{Q^{\prime}})[n]\rangle+\bar{Q}Q^{\prime}$ ($Q^{(\prime)}=c~\text{or}~b$ quark) within the NRQCD framework. In addition to the lower-level Fock states $|(Q\bar{Q'})[1S]\rangle$ continent, contributions of high excited states $|(Q\bar{Q'})[2S]\rangle$, $|(Q\bar{Q'})[3S]\rangle$, $|(Q\bar{Q'})[4S]\rangle$, $|(Q\bar{Q'})[1P]\rangle$, $|(Q\bar{Q'})[2P]\rangle$, $|(Q\bar{Q'})[3P]\rangle$ and $|(Q\bar{Q'})[4P]\rangle$ are also studied. According to our study, the contributions of high excited Fock states should be considered seriously. Differential distributions of total decay width with respect to invariant-mass and angles, as well as uncertainties caused by non-perturbative hadronic non-perturbative matrix elements are discussed.  If all excited heavy quarkonium states decay to the ground spin-singlet state through electromagnetic or hadronic interactions, we obtain the decay widths for $|(Q\bar{Q'})\rangle$ quarkonium production through $H^0$ semi-exclusive decays: $25.10^{+11.6\%}_{-51.6\%}$ keV for $|(b\bar{c})[n]\rangle$ meson, $3.23^{+0\%}_{-62.2\%}$ keV for $|(c\bar{c})[n]\rangle$ and $2.36^{+0\%}_{-57.1\%}$ keV for $|(b\bar{b})[n]\rangle$, where uncertainties are caused by adopting different non-perturbative potential models. At future high energy LHC ($\sqrt{s}=27$ TeV), numerical results show that sizable amounts of events for those high excited states can be produced, which implies that one could also consider exploring the coupling properties of Higgs to heavy quarks in these high excited states channels, especially for the charmonium and bottomonium.\\


\end{abstract}

\maketitle

\section{Introduction}

The Higgs boson of the Standard Model (SM) has been found and confirmed by the CMS and ATLAS collaborations at the Large Hadron Collider (LHC) \cite{cms1,atlas1,cms2,atlas2}. However to reveal the nature of Higgs boson, we need to further study its coupling to fundamental particles as well as the Higgs self-coupling interaction. Though the LHC has made great progress in understanding the coupling properties of Higgs to vector bosons and heavy fermions \cite{Unal:2018tyf}, the measuring precision are restricted due to the limited dataset and complicated hadronic background.
There are two major upgrade of the LHC, i.e., high luminosity/energy LHC (HL/HE-LHC), which provide excellent opportunities in Higgs physics \cite{lhc}. Precise measurements can also be performed in the clean environment of the future electron-positron colliders, like the Circular Electron-Positron Collider (CEPC) \cite{cepc} and the International Linear Collider (ILC) \cite{hb}.

The CMS and ATLAS collaborations have reported measurements of the Higgs boson coupling to the third-generation fermions, i.e., $H^0\to\tau^+\tau^-$ decay \cite{Khachatryan:2016vau}, associated production of Higgs boson with a top pair \cite{Sirunyan:2018hoz,Aaboud:2018urx} and the $H^0\to b\bar{b}$ channel \cite{Aaboud:2017xsd,Sirunyan:2017elk}. But no evidence of Higgs boson coupling to the first- and second-generation fermions, except for the direct searches for $H^0\to c\bar{c}$ \cite{Aaboud:2018fhh}, $H^0\to \mu^+\mu^-$ and $H^0\to e^+e^-$ \cite{Aaboud:2017ojs,Khachatryan:2014aep}. As a manner complementary to studies of the direct exploration, the heavy quarkonium production in Higgs boson decays might also be taking into consideration seriously in the proposed  HL/HE-LHC, CEPC or ILC platforms.
Continuous exploration on the search for $H^0\rightarrow J/\Psi \gamma$ and $H^0\rightarrow \Upsilon(nS) \gamma$ have been carried out by ATLAS \cite{Aad:2015sda,Aaboud:2018txb}. The former decay mode is also explored  at CMS \cite{Khachatryan:2015lga}.

Theoretically, related calculations have been studied \cite{nv,cfk,Bodwin:2013gca,Bodwin:2014bpa,gf1,gf2,Koenig:2015pha,Brambilla:2019fmu}. Within the nonrelativistic quantum chromodynamics (NRQCD) formalism \cite{nrqcd1,nrqcd2} and light-cone methods \cite{gs,Chernyak:1983ej}, both the direct and loop-induced indirect production mechanism \cite{Bodwin:2013gca,Bodwin:2014bpa,gf1,gf2,Koenig:2015pha,Brambilla:2019fmu}, the relativistic corrections  \cite{Bodwin:2014bpa,Brambilla:2019fmu} and the resummation contributions \cite{gf1,gf2} to $H^0\rightarrow J/\Psi \gamma$ and $H^0\rightarrow \Upsilon(nS) \gamma$ are studied.
The semi-exclusive $B^{(*)}_c$ meson production in Higgs boson decays, $H^0\to B^{(*)}_c+\bar{c}b$ is also systematically investigated \cite{jjcq}. It is found that the $Hb\bar{b}$ coupling Feynman diagrams dominate the process, while contributions from the triangle top-quark loop,  $Hc\bar{c},~HWW$ and $HZZ$ coupling diagrams are comparatively negligible.
Interestingly, we also find that decay width of Higgs boson to $B^{(*)}_c$ meson is larger than those of Higgs to charmonium and bottomonium by almost an order of magnitude.
Moreover, the production of $|(Q\bar{Q'})\rangle$ ($Q^(\prime)=c,~b$) quarkonium through the color-octet (c.o.) Fock states configuration is much smaller than those through the color-singlet (c.s.) configuration \cite{lx1}, for example $\Gamma(H^0\to |(c\bar{b})[1S]\rangle+\bar{c}b)_{c.o.}/\Gamma(H^0\to |(c\bar{b})[1S]\rangle+\bar{c}b)_{c.s.} \simeq 0.005$.

According to our study, the high excited quarkonium states, i.e., $n^1S_0, ~n^3S_1, ~n^1P_1, ~n^3P_J$ $(n=1,2,3,4, J=0,1,2)$ can also be generated massively in comparison with the ground state $1^1S_0$ at future HL/HE-LHC \cite{Liao:2012rh,lx,Liao:2015vqa}. Here to illustrate this issue, we present some examples. In the $W^{+}\rightarrow |(c\bar{c})[n]\rangle + c \bar{s}$ channel, the decay widths for $[n]=2S$, $3S$, $1P$ and $2P$ $|(c\bar{c})[n]\rangle$ states are about $43\%$, $21\%$, $35\%$ and $21\%$ of that of the $1S$ configuration \cite{Liao:2012rh}. For $|(b\bar{b})[n]\rangle$ quarkonium production in $t\rightarrow |(b\bar{b})[n]\rangle+ bW^{+}$ process, the total decay widths for $2S$, $3S$, $1P$, $2P$, wave states are about $31.9\%$, $9.2\%$, $15.0\%$ and $6.0\%$ of those of $1S$ bottomonium \cite{lx}. And in $Z^0\rightarrow |(b\bar{c})[n]\rangle+\bar{b}c$ channel, total decay widths for $2S$, $3S$, $1P$, $2P$ Fock states are $24.8\%$, $13.3\%$, $8.5\%$, $4.7\%$ of the summed decay widths of $B_c$ and $B^*_c$ \cite{Liao:2015vqa}. Here $nS$ ($n=1, 2, 3,4$) stands for the summed decay widths of $n^1S_0$ and $n^3S_1$ at the same $n$th level, and $nP$ stands for the summed decay width of $n^1P_1$ and $n^3P_J$ ($J=0,1,2$) at the same $n$th level. Numerical results show that excited $nS$ and $nP$ wave states can provide sizable contributions to heavy quarkonium production, which implies that one might explore the coupling properties of Higgs boson to heavy quarks using the dataset of these high excited quarkonium production channels.

In our previous work \cite{lx1} , we study the P-wave and color-octet configuration quarkonium production in Higgs semi-exlusive decays under the NRQCD factorization framework. In this manuscript, we further study the production of high excited Fock states of $|(b\bar{c})[n]\rangle$, $|(c\bar{c})[n]\rangle$ and $|(b\bar{b})[n]\rangle$ quarkonia in Higgs boson decays, i.e., $n^1S_0$, $n^3S_1$, $n^1P_0$ and $n^3P_J$ ($n=1,2,3,4$; $J=0, 1, 2$) configuration.  We believe that, to derive more precise coupling parameters of Higgs to fermions in Higgs to heavy quarkonia rare decays, contributions from these high excited states together with uncertainties caused by the non-perturbative parameters, Higgs and quark masses should be seriously discussed.

As is known that analytical expressions for the usual squared amplitudes become complex and lengthy for massive particles in the final states especially to derive the amplitudes of the $P$-wave Fock states. To solve this problem, the ``improved trace technology" is suggested and developed \cite{tbc2,zbc1,wbc1}, which is based on the helicity amplitudes method and deals with the trace calculation directly at the amplitude level. In this paper, we continue to adopt ``improved trace technology" to derive the analytical expression for all the decay channels.

The rest of the present work is organized as follows. In Section II, we introduce the calculation formalism for the $H^0$ boson semi-exclusive decays to $|(Q\bar{Q'})[n]\rangle$ $(Q^{(\prime)}=c~\text{or}~b)$ quarkonium within the NRQCD framework. In Section III, we evaluate the decay widths of $H^0\to |(b\bar{c})[n]\rangle+\bar{b}c$, $H^0\to |(c\bar{c})[n]\rangle+\bar{c}c$ and $H^0\to |(b\bar{b})[n]\rangle+\bar{b}b$, where $[n]$ stands for $n^1S_0$, $n^3S_1$, $n^1P_0$ and $n^3P_J$ ($n=1,2,3,4$; $J=0, 1, 2$). To further illustrate contributions of the high excited Fock states, differential distributions of decay widths with respect to invariant-mass and angles, as well as uncertainties caused by non-perturbative hadronic parameters under five different potential models, are studied in detail. We also present an estimation on the total heavy quarkonium events at the proposed HE-LHC. The final Section IV is reserved for a summary.

\section{Calculation Techniques and Formulations}

The semi-exclusive decay processes of Higgs boson to heavy quarkonia, i.e., $H^0\to |(c\bar{b})[n]\rangle+\bar{c}b$ (or $H^0\to |(b\bar{c})[n]\rangle+\bar{b}c$), $H^0\to |(c\bar{c})[n]\rangle+\bar{c}c$ and $H^0\to |(b\bar{b})[n]\rangle+\bar{b}b$, can be dealt with analogously within the NRQCD factorization framework \cite{nrqcd1,nrqcd2}. Here the squared amplitudes can be factoried as the production of the perturbatively calculable short-distance coefficients and the non-perturbative long-distance factors, the so-called non-perturbative NRQCD matrix elements. The total decay widths $d\Gamma$ can be written as
\begin{equation}
d\Gamma=\sum_{n}{\langle{\cal O}^H(n) \rangle} d\hat\Gamma(H^0\to |(Q\bar{Q'})[n]\rangle+ \bar{Q'}Q).
\end{equation}
Here $\langle{\cal O}^{H}(n)\rangle$ is the non-perturbative matrix element, which describes the hadronization of a $|(Q\bar{Q'})[n]\rangle$ Fock state into the observable heavy quarkonium. For the color-singlet Fock states, the non-perturbative matrix elements can be directly related either to the wave functions at the origin for $nS$-wave states, or to the first derivative of the wave functions at the origin for $nP$-wave states \cite{nrqcd1}, which can be calculated through the potential NRQCD \cite{pnrqcd1,yellow}, lattice QCD \cite{lat1}, or the potential models \cite{lx,Eichten:1978tg,Eichten:1979ms,Eichten:1980mw,pot2,pot3,pot4,Chen:1992fq}.

\begin{figure}
\includegraphics[width=0.45\textwidth]{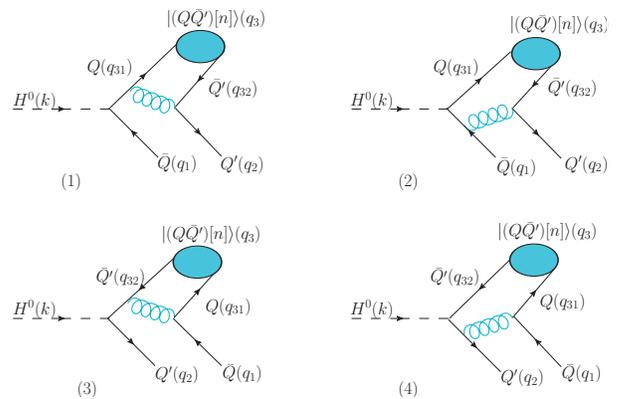}
\caption{(color online). Feynman diagrams for processes of $H^0(k)\rightarrow |(Q\bar{Q'})[n]\rangle(q_3) + Q'(q_2)+\bar{Q}(q_1) ~(Q^{(\prime)}=c~\text{or}~b~\text{quark})$, where $|(Q\bar{Q'})[n]\rangle$ stands for $|(b\bar{c})[n]\rangle$, $|(c\bar{c})[n]\rangle$ and $|(b\bar{b})[n]\rangle$ quarkonia. Here $[n]$ is short for $[n^1S_0]$, $[n^3S_1]$, $[n^1P_0]$ and $[n^3P_J]$ Fock states with $n=1,2,3,4, J=0,1,2$.} \label{feyn1}
\end{figure}

The short-distance decay width $\hat\Gamma$ can be expressed as
\begin{equation}
d\hat\Gamma(H^0\to |(Q\bar{Q'})[n]\rangle+\bar{Q}Q')= \frac{1}{2 m_{H}} \overline{\sum}  |{\cal M}(n)|^{2} d\Phi_3,
\end{equation}
where $m_H$ is the mass of the Higgs boson, $\overline{\sum}$ means that one needs to average over the spin states of the initial particles and to sum over the color and spin of all the final particles when manipulating the squared amplitudes $|M(n)|^{2}$. In the Higgs boson rest frame, the three-particle phase space can be written as
\begin{equation}
d{\Phi_3}=(2\pi)^4 \delta^{4}\left(k_{0} - \sum_f^3 q_{f}\right)\prod_{f=1}^3 \frac{d^3{\vec{q}_f}}{(2\pi)^3 2q_f^0}.
\end{equation}
With the help of the formulas listed in Refs.~\cite{tbc2,zbc1}, one can also derive the corresponding differential decay widths that are helpful for experimental studies, i.e., $d\Gamma/ds_{1}$, $d\Gamma/ds_{2}$, $d\Gamma/d\cos\theta_{12}$ and $d\Gamma/d\cos\theta_{13}$, where $s_{1}=(q_1+q_2)^2$, $s_{2}=(q_1+q_3)^2$, $\theta_{12}$ is the angle between $\vec{q}_1$ and $\vec{q}_2$, and $\theta_{23}$ is that between $\vec{q}_2$ and $\vec{q}_3$. We will discuss these differential distributions during the numerical estimation.

To further illustrate the above processes of $H^0(k) \rightarrow |(Q\bar{Q'})[n]\rangle(q_3) +Q'(q_2)+ \bar{Q}(q_1)~(Q^{(\prime)}=c~\text{or}~b~\text{quark})$ ,  the Feynman diagrams are presented in Fig.~\ref{feyn1}. And the general form of their amplitudes can be expressed as
\begin{equation} \label{amplitude}
i{\cal M}(n)= {\cal{C}} {\bar {u}_{s i}}({q_2}) \sum\limits_{k = 1}^{4} {{\cal A} _k } {v_{s' j}}({q_1}),
\end{equation}
where the overall factor ${\cal C}=g_s^2\frac{C_{F}\delta_{ij}}{\sqrt{N_c}}\frac{g}{2 m_{W}}$ with $N_c=3$ for the QCD, $k$ is the number of Feynman diagrams, $s$ and $s'$ are spin states, $i$ and $j$ are color indices for the $Q$-quark and $\bar{Q}$-quark, and ${\cal A} _k $ is the Dirac matrix chain.

The explicit expressions of ${\cal A}_k$ for the $nS$-wave states ($n=1,2,3,4$) can be written as
\begin{eqnarray}
{\cal A}_1 &=& \left[{m_Q}{\gamma_\alpha} \frac{\Pi^{0(\nu)}_{q_3}(q)}{(q_2 + {q_{32}})^2} {\gamma_\alpha} \frac{(\slashed{q}_2+\slashed{q}_3)+{m_Q}}{(q_2+q_3 )^2 - m_Q^2} \right]_{q=0}, \label{A1}\\
{\cal A}_2 &=& \left[{m_Q}{\gamma_\alpha} \frac{\Pi^{0(\nu)}_{q_3}(q)}{(q_2 + {q_{32}})^2} \frac{-(\slashed{k}-\slashed{q}_{31})+{m_Q}}{(k-q_{31} )^2 - m_Q^2}{\gamma_\alpha} \right]_{q=0}, \label{A2}\\
{\cal A}_3 &=& \left[{m_{Q'}} \frac{-(\slashed{q_1}+\slashed{q}_{3})+{m_{Q'}}}{(q_1+q_{3} )^2 - m_{Q'}^2}{\gamma_\alpha} \frac{\Pi^{0(\nu)}_{q_3}(q)}{(q_1 + {q_{31}})^2} {\gamma_\alpha} \right]_{q=0}, \label{A3}\\
{\cal A}_4 &=&  \left[{m_{Q'}} {\gamma_\alpha} \frac{(\slashed{k}-\slashed{q}_{32})+{m_{Q'}}}{(k-q_{32} )^2 - m_{Q'}^2} \frac{\Pi^{0(\nu)}_{q_3}(q)}{(q_1 + {q_{31}})^2} {\gamma_\alpha} \right]_{q=0}. \label{A4}
\end{eqnarray}
Here $\Pi^{0(\nu)}(q)$ are the projectors for spin-singlet (spin-triplet) states, $q$ stands for the relative momentum between the two constituent quarks in the $|(Q\bar{Q'})[n]\rangle$ state,
\begin{eqnarray}
\Pi^0_{q_3}(q)&=&\frac{-\sqrt{m_{Q\bar{Q'}}}}{4{m_Q}{m_{Q'}}}(\slashed{q}_{32}- m_{Q'}) \gamma_5 (\slashed{q}_{31} + m_Q),\\
\Pi^\nu_{q_3}(q)&=&\frac{-\sqrt{m_{Q\bar{Q'}}}}{4{m_Q}{m_{Q'}}}(\slashed{q}_{32}- m_{Q'}) \gamma_\nu (\slashed{q}_{31} + m_Q).
\end{eqnarray}
$q_{31}$ and $q_{32}$ are the momenta of the two constituent quarks,
\begin{eqnarray}
q_{31}&=&\frac{{m_b}}{m_{Q\bar{Q'}}}q_3+q,\\
q_{32}&=&\frac{{m_Q}}{m_{Q\bar{Q'}}}q_3-q,
\end{eqnarray}
where $m_{Q\bar{Q'}}=m_Q+m_{Q'}$ is implicitly adopted.

For the $n^1P_1$-wave states ($n=1,2,3,4$), ${\cal A}_k$ can be written as
\begin{widetext}
\begin{eqnarray}
{\cal A}^{S=0,L=1}_1 &=& \varepsilon_l^{\mu}(q_3) \frac{d}{dq_\mu} \left[{m_Q}{\gamma_\alpha} \frac{\Pi^{0}_{q_3}(q)}{(q_2 + {q_{32}})^2} {\gamma_\alpha} \frac{(\slashed{q}_2+\slashed{q}_3)+{m_Q}}{(q_2+q_3 )^2 - m_Q^2} \right]_{q=0}, \label{A5}\\
{\cal A}^{S=0,L=1}_2 &=& \varepsilon_l^{\mu}(q_3) \frac{d}{dq_\mu} \left[{m_Q}{\gamma_\alpha} \frac{\Pi^{0}_{q_3}(q)}{(q_2 + {q_{32}})^2} \frac{-(\slashed{k}-\slashed{q}_{31})+{m_Q}}{(k-q_{31} )^2 - m_Q^2}{\gamma_\alpha} \right]_{q=0}, \label{A6}\\
{\cal A}^{S=0,L=1}_3 &=& \varepsilon_l^{\mu}(q_3) \frac{d}{dq_\mu} \left[{m_{Q'}} \frac{-(\slashed{q_1}+\slashed{q}_{3})+{m_{Q'}}}{(q_1+q_{3} )^2 - m_{Q'}^2}{\gamma_\alpha} \frac{\Pi^{0}_{q_3}(q)}{(q_1 + {q_{31}})^2} {\gamma_\alpha} \right]_{q=0}, \label{A7}\\
{\cal A}^{S=0,L=1}_4 &=& \varepsilon_l^{\mu}(q_3) \frac{d}{dq_\mu} \left[{m_{Q'}} {\gamma_\alpha} \frac{(\slashed{k}-\slashed{q}_{32})+{m_{Q'}}}{(k-q_{32} )^2 - m_{Q'}^2} \frac{\Pi^{0}_{q_3}(q)}{(q_1 + {q_{31}})^2} {\gamma_\alpha} \right]_{q=0}, \label{A8}
\end{eqnarray}
where $\varepsilon_{l}^{\mu}(q_3)$ are the polarization vectors relating to the orbit angular momentum of the $|(Q\bar{Q'})[n^1P_1]\rangle$ state.
And for the $n^3P_J$-wave states ($n=1,2,3,4$; $J=0,1,2$),
\begin{eqnarray}
{\cal A}^{S=1,L=1}_1 &=& \varepsilon^{J}_{\mu\nu}(q_3) \frac{d}{dq_\mu}  \left[{m_Q}{\gamma_\alpha} \frac{\Pi^{\nu}_{q_3}(q)}{(q_2 + {q_{32}})^2} {\gamma_\alpha} \frac{(\slashed{q}_2+\slashed{q}_3)+{m_Q}}{(q_2+q_3 )^2 - m_Q^2} \right]_{q=0}, \label{A9}\\
{\cal A}^{S=1,L=1}_2 &=& \varepsilon^{J}_{\mu\nu}(q_3) \frac{d}{dq_\mu} \left[{m_Q}{\gamma_\alpha} \frac{\Pi^{\nu}_{q_3}(q)}{(q_2 + {q_{32}})^2} \frac{-(\slashed{k}-\slashed{q}_{31})+{m_Q}}{(k-q_{31} )^2 - m_Q^2}{\gamma_\alpha} \right]_{q=0}, \label{A10}\\
{\cal A}^{S=1,L=1}_3 &=& \varepsilon^{J}_{\mu\nu}(q_3) \frac{d}{dq_\mu}  \left[{m_{Q'}} \frac{-(\slashed{q_1}+\slashed{q}_{3})+{m_{Q'}}}{(q_1+q_{3} )^2 - m_{Q'}^2}{\gamma_\alpha} \frac{\Pi^{\nu}_{q_3}(q)}{(q_1 + {q_{31}})^2} {\gamma_\alpha} \right]_{q=0}, \label{A11}\\
{\cal A}^{S=1,L=1}_4 &=& \varepsilon^{J}_{\mu\nu}(q_3) \frac{d}{dq_\mu} \left[{m_{Q'}} {\gamma_\alpha} \frac{(\slashed{k}-\slashed{q}_{32})+{m_{Q'}}}{(k-q_{32} )^2 - m_{Q'}^2} \frac{\Pi^{\nu}_{q_3}(q)}{(q_1 + {q_{31}})^2} {\gamma_\alpha} \right]_{q=0}, \label{A12}
\end{eqnarray}
where $\varepsilon^{J}_{\mu\nu}(q_3)$ is the polarization tensor for the spin triplet $P$-wave states with $J=0,1,2$.
\end{widetext}

Selection of the appropriate total angular momentum quantum number is done by performing the proper polarization sum. For a spin-triplet $S$ state or a spin-singlet $P$ state, it is given by \cite{projector}
\begin{eqnarray}\label{3pja}
\sum_{J_z}\varepsilon^{0}_{\alpha} \varepsilon^{0*}_{\beta} = \Pi_{\alpha\beta} = -g_{\alpha\beta}+\frac{q_{3\alpha}q_{3\beta}}{m_{Q\bar{Q'}}^{2}},
\end{eqnarray}
where $J_z=s_z$ or $l_z$ respectively. In the case of $^3P_J$ states, the sum over the polarization is given by \cite{projector}
\begin{eqnarray}\label{3pja}
\sum_{J_z}\varepsilon^{0}_{\mu\nu} \varepsilon^{0*}_{\mu'\nu'}&=& \frac{1}{3} \Pi_{\mu\nu}\Pi_{\mu'\nu'}, \\
\sum_{J_z}\varepsilon^{1}_{\mu\nu} \varepsilon^{1*}_{\mu'\nu'}&=& \frac{1}{2}
(\Pi_{\mu\mu'}\Pi_{\nu\nu'}- \Pi_{\mu\nu'}\Pi_{\mu'\nu}) \label{3pjb},\\
\sum_{J_z}\varepsilon^{2}_{\mu\nu} \varepsilon^{2*}_{\mu'\nu'}&=& \frac{1}{2}
(\Pi_{\mu\mu'}\Pi_{\nu\nu'}+ \Pi_{\mu\nu'}\Pi_{\mu'\nu})-\frac{1}{3} \Pi_{\mu\nu}\Pi_{\mu'\nu'}, \nonumber \\ \label{3pjc}
\end{eqnarray}
for $J=0,1,2$, respectively.

To improve the efficiency of numerical evaluation, we adopt the ``improved trace technology" to simplify the amplitudes ${\cal M}(n)$ at the amplitude level. To shorten the manuscript, we will not repeat the derivation process here. For technical details and examples, one can refer to literatures \cite{tbc2,zbc1,wbc1}.

In our formalism, the main uncertainty would be from the color-singlet non-perturbative matrix element $\langle{\cal O}^H(n) \rangle$, which can be related to the Schr\"{o}dinger wave function at the origin $\psi_{(Q\bar{Q'})}(0)$ for the $nS$-wave Fock states or the first derivative of the wave function at the origin $\psi^\prime_{(Q\bar{Q'})}(0)$ for the $nP$-wave states:
\begin{eqnarray}
\langle{\cal O}^H(nS) \rangle &\simeq& |\Psi_{\mid(Q\bar{Q'})[nS]\rangle}(0)|^2,\nonumber\\
\langle{\cal O}^H(nP) \rangle &\simeq& |\Psi^\prime_{\mid(Q\bar{Q'})[nP]\rangle}(0)|^2.
\end{eqnarray}
Due to the fact that spin-splitting effects are small at the same level, we adopt the same wave function values for both the spin-singlet and spin-triplet states here. Further, the Schr\"{o}dinger wave function at the origin $\Psi_{|Q\bar{Q'})[nS]\rangle}(0)$ and its first derivative $\Psi^{'}_{|(Q\bar{Q'})[nP]\rangle}(0)$ are related to the radial wave function at the origin $R_{|(Q\bar{Q'})[nS]\rangle}(0)$ and its first derivative $R^{'}_{|(Q\bar{Q'})[nP]\rangle}(0)$, respectively~\cite{nrqcd1,lx}:
\begin{eqnarray}
\Psi_{|(Q\bar{Q'})[nS]\rangle}(0)&=&\sqrt{{1}/{4\pi}}R_{|(Q\bar{Q'})[nS]\rangle}(0),\nonumber\\
\Psi'_{|(Q\bar{Q'})[nP]\rangle}(0)&=&\sqrt{{3}/{4\pi}}R'_{|(Q\bar{Q'})[nP]\rangle}(0).
\end{eqnarray}
In the manuscript of Ref.~\cite{lx}, we present a systematic study on these radial wave function at the origin $R_{|(Q\bar{Q'})[nS]\rangle}(0)$ for $nS$-wave quarkonium states, the first derivative $R^{'}_{|(Q\bar{Q'})[nP]\rangle}(0)$ for $nP$-wave states and the second derivative $R^{''}_{|(Q\bar{Q'})[nD]\rangle}(0)$ for $nD$-wave states under five different potential models. In Section III (C), we will discuss the uncertainties of the decay widths of $\Gamma(H^0\to |(Q\bar{Q'})[n]\rangle+ \bar{Q'}Q)$ ($Q^{(\prime)}=c~\text{or}~b$ quark)  caused by these radial wave functions in detail.

\section{Numerical Results}

\subsection{Input parameters}

In the numerical computation, we adopt the running strong coupling parameter $\alpha_s$, i.e., $\alpha_s=0.26$ for $|(c\bar{c})\rangle$ and $|(b\bar{c})\rangle$-quarkonia, and $\alpha_s=0.18$ for $|(b\bar{b})\rangle$-quarkonium.
Because the Buchm\"{u}ller and Tye potential model (BT-potential) has the correct two-loop short-distance behavior in QCD~\cite{pot2,wgs}, wave functions evaluated under the BT-potential are adopted. Specifically, one can find values of the radial wave functions at the origin, and the first derivative of the radial wave functions at the origin for the $|(Q\bar{Q'})[n]\rangle$ ($Q^{(\prime)}=c~\text{or}~b$ quark) quarkonia in tables I, II and III in our earlier manuscript~\cite{lx}. To shorten this manuscript, we do not present them here. Other parameters are adopted as the following values \cite{pdg}: $m_c=1.45$ GeV, $m_b=4.85$ GeV, $m_H =125.18$ GeV, the Fermi constant $G_F=1.16639$ GeV$^{-2}$, the Weinberg angle $\theta_W=\arcsin\sqrt{0.23119}$, and the total decay width of Higgs boson $\Gamma_{H^0}=4.2$ MeV \cite{sh}.  To ensure the gauge invariance of the hard amplitude, we set the $|(Q\bar{Q'})[n]\rangle$ quarkonium mass $m_{Q\bar{Q'}}$ to be $m_Q+m_{Q'}$.

\subsection{Heavy quarkonium production in $H^0$ decays}

\begin{table}
\caption{Decay widths (units: $keV$) for the production of high excited states $|(b\bar{c})[n]\rangle$ quarkonium through Higgs boson decays within the BT-potential model ($n_f=4$)~\cite{lx}.}
\begin{tabular}{|c|c|c|c|c|c|}
\hline
$H^0\to |(b\bar{c})[n]\rangle+\bar{b}c$&~$n=1$~&~$n=2$~&~$n=3$~&~$n=4$~\\
\hline\hline
$\Gamma(H^0\to |(b\bar{c})[n^1S_0]\rangle +\bar{b}c)$&~5.736~&~1.135~&~0.8251~&~0.7619~\\
\hline
$\Gamma(H^0\to |(b\bar{c})[n^3S_1]\rangle +\bar{b}c)$&~7.857~&~1.445~&~1.028~&~0.9317~\\
\hline
$\Gamma(H^0\to |(b\bar{c})[n^1P_1]\rangle +\bar{b}c)$&~0.2761~&~0.1478~&~0.1740~&~0.1710~\\
\hline
$\Gamma(H^0\to |(b\bar{c})[n^3P_0]\rangle +\bar{b}c)$&~0.1838~&~0.1031~&~0.1297~&~0.1315~\\
\hline
$\Gamma(H^0\to |(b\bar{c})[n^3P_1]\rangle +\bar{b}c)$&~0.6706~&~0.3517~&~0.4176~&~0.4098~\\
\hline
$\Gamma(H^0\to |(b\bar{c})[n^3P_2]\rangle +\bar{b}c)$&~0.3521~&~0.1763~&~0.2001~&~0.1946~\\
\hline
Sum&~15.08~&~3.359~&~2.775~&~2.601~\\
\hline\hline
\end{tabular}
\label{tabrpa}
\end{table}
\begin{table}
\caption{Decay widths (units: $eV$) for the production of high excited states $|(c\bar{c})[n]\rangle$ quarkonium through Higgs boson decays within the BT-potential model ($n_f=4$)~\cite{lx}.}
\begin{tabular}{|c|c|c|c|c|c|}
\hline
$H^0\to |(c\bar{c})[n]\rangle+\bar{c}c$&~$n=1$~&~$n=2$~&~$n=3$~&~$n=4$~\\
\hline\hline
$\Gamma(H^0\to |(c\bar{c})[n^1S_0]\rangle +\bar{c}c)$&~616.6~&~293.2~&~180.7~&~154.8~\\
\hline
$\Gamma(H^0\to |(c\bar{c})[n^3S_1]\rangle +\bar{c}c)$&~594.8~&~276.4~&~169.1~&~143.9~\\
\hline
$\Gamma(H^0\to |(c\bar{c})[n^1P_1]\rangle +\bar{c}c)$&~70.81~&~35.06~&~44.57~&~45.52~\\
\hline
$\Gamma(H^0\to |(c\bar{c})[n^3P_0]\rangle +\bar{c}c)$&~104.5~&~51.73~&~~67.02~&~69.65~\\
\hline
$\Gamma(H^0\to |(c\bar{c})[n^3P_1]\rangle +\bar{c}c)$&~66.46~&~32.90~&~42.54~&~43.85~\\
\hline
$\Gamma(H^0\to |(c\bar{c})[n^3P_2]\rangle +\bar{c}c)$&~45.04~&~21.88~&~28.93~&~29.68~\\
\hline
Sum&~1498~&~711.2~&~532.9~&~487.4~\\
\hline\hline
\end{tabular}
\label{tabrpb}
\end{table}
\begin{table}
\caption{Decay widths (units: $eV$) for the production of high excited states $|(b\bar{b})[n]\rangle$ quarkonium through Higgs boson decays within the BT-potential model ($n_f=5$)~\cite{lx}.}
\begin{tabular}{|c|c|c|c|c|c|}
\hline
$H^0\to |(b\bar{b})[n]\rangle+\bar{b}b$&~$n=1$~&~$n=2$~&~$n=3$~&~$n=4$~\\
\hline\hline
$\Gamma(H^0\to |(b\bar{b})[n^1S_0]\rangle +\bar{b}b)$&~591.1~&~295.6~&~187.0~&~112.4~\\
\hline
$\Gamma(H^0\to |(b\bar{b})[n^3S_1]\rangle +\bar{b}b)$&~445.9~&~217.7~&~136.0~&~81.15~\\
\hline
$\Gamma(H^0\to |(b\bar{b})[n^1P_1]\rangle +\bar{b}b)$&~18.08~&~16.78~&~12.25~&~8.266~\\
\hline
$\Gamma(H^0\to |(b\bar{b})[n^3P_0]\rangle +\bar{b}b)$&~39.83~&~31.75~&~23.32~&~15.95~\\
\hline
$\Gamma(H^0\to |(b\bar{b})[n^3P_1]\rangle +\bar{b}b)$&~32.79~&~25.57~&~17.18~&~9.588~\\
\hline
$\Gamma(H^0\to |(b\bar{b})[n^3P_2]\rangle +\bar{b}b)$&~13.23~&~12.37~&~9.055~&~6.145~\\
\hline
Sum&~1141~&~599.8~&~384.8~&~233.5~\\
\hline\hline
\end{tabular}
\label{tabrpc}
\end{table}

\begin{figure}[htbp]
\centering
\includegraphics[width=0.23\textwidth]{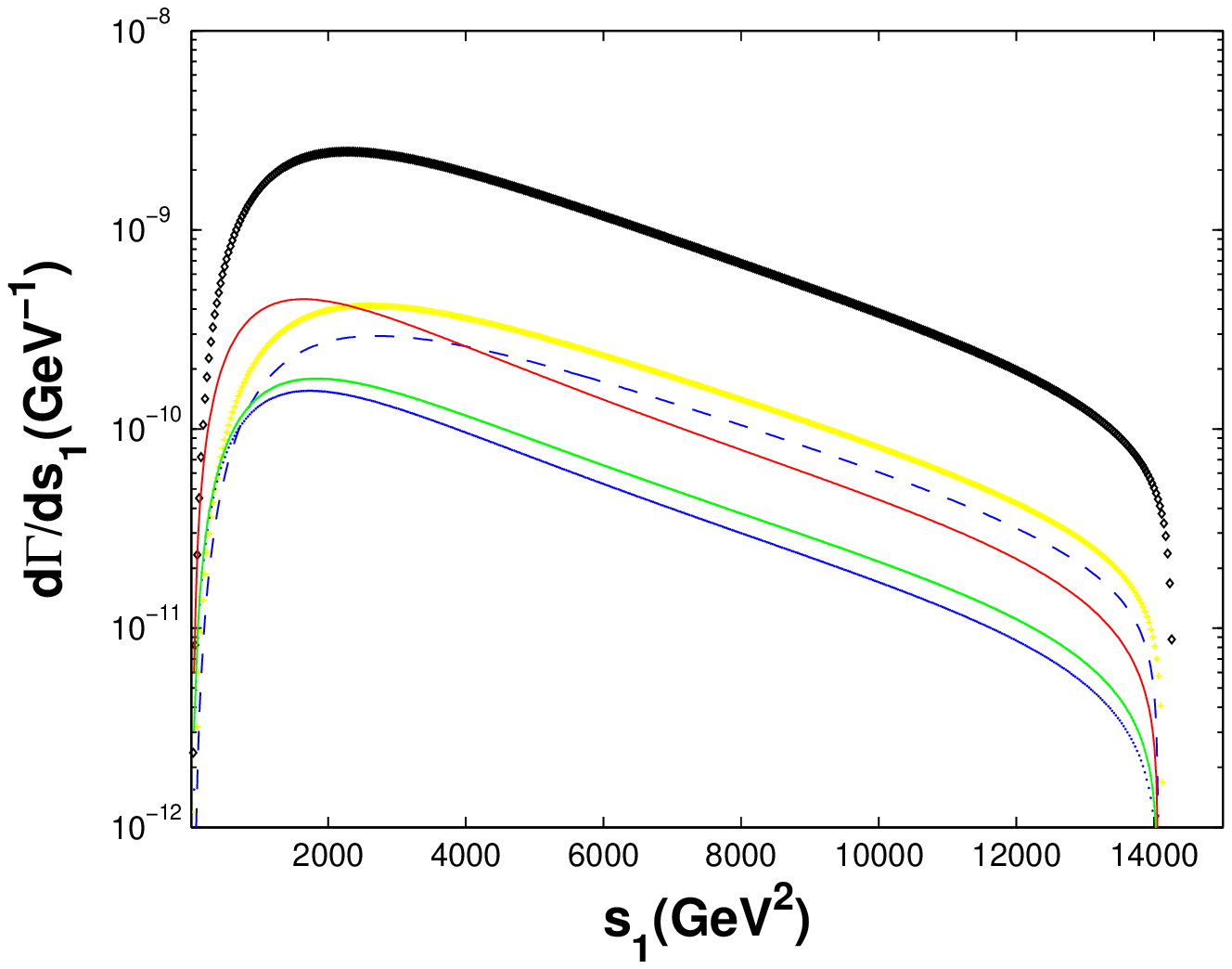}
\includegraphics[width=0.23\textwidth]{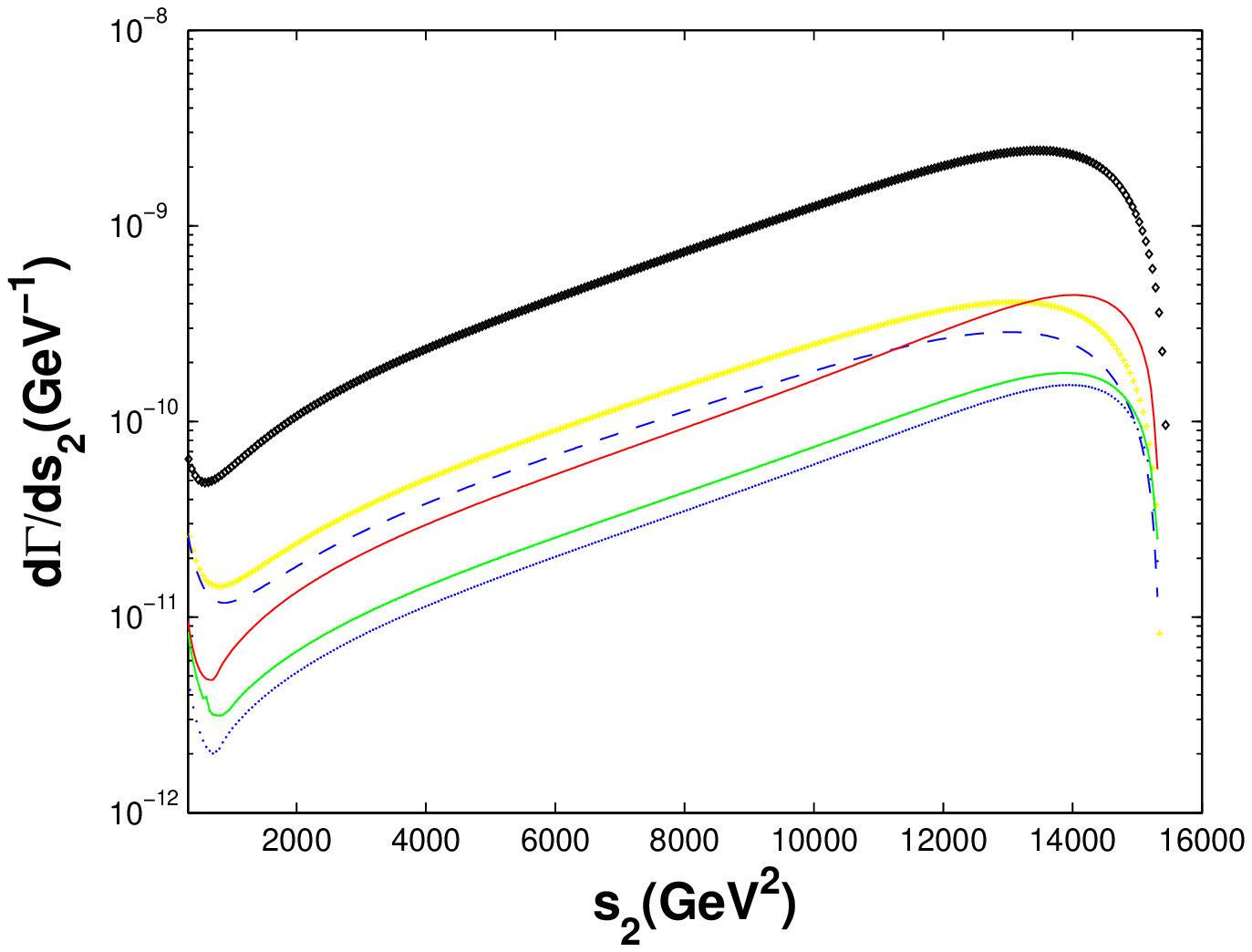}
\includegraphics[width=0.23\textwidth]{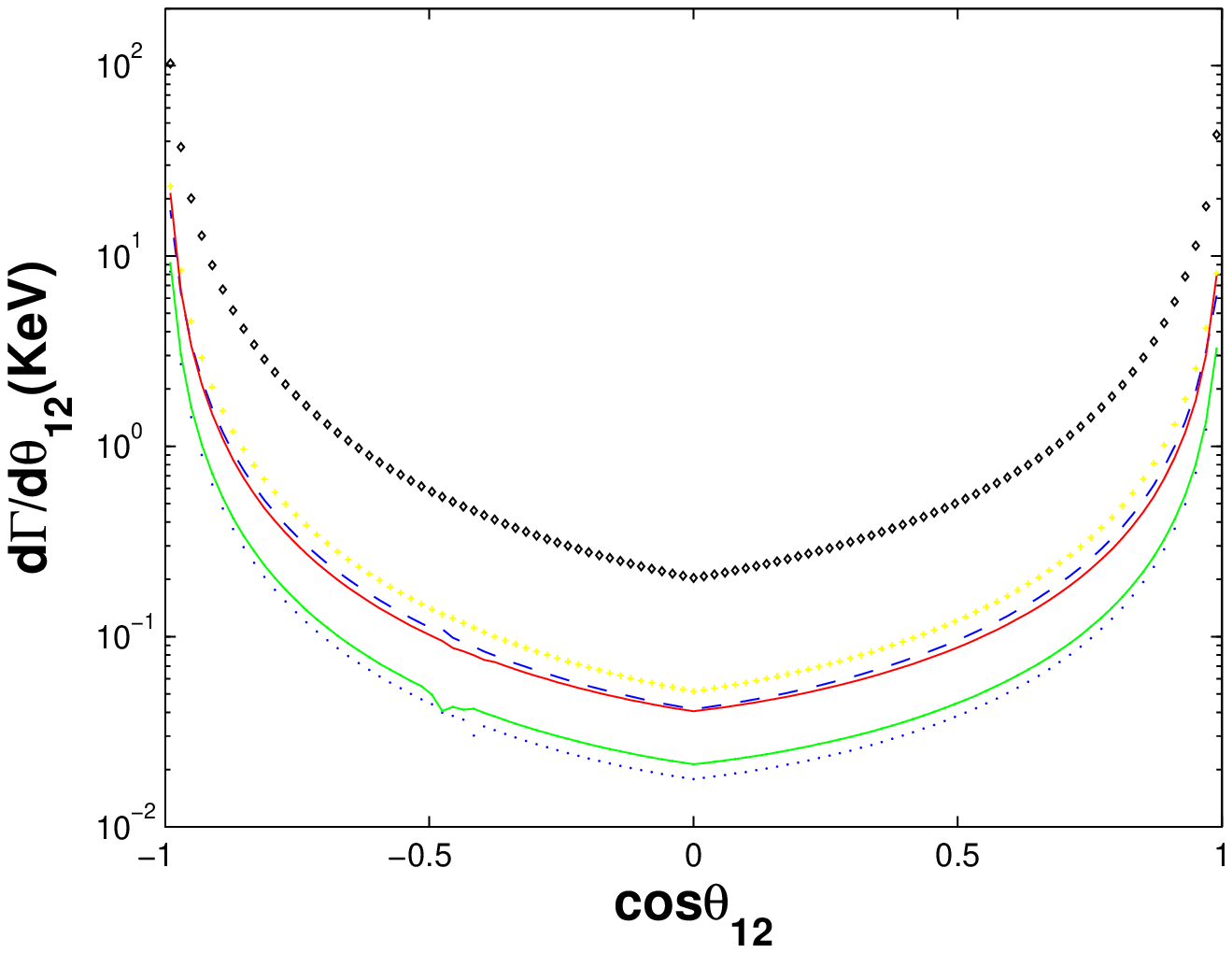}
\includegraphics[width=0.23\textwidth]{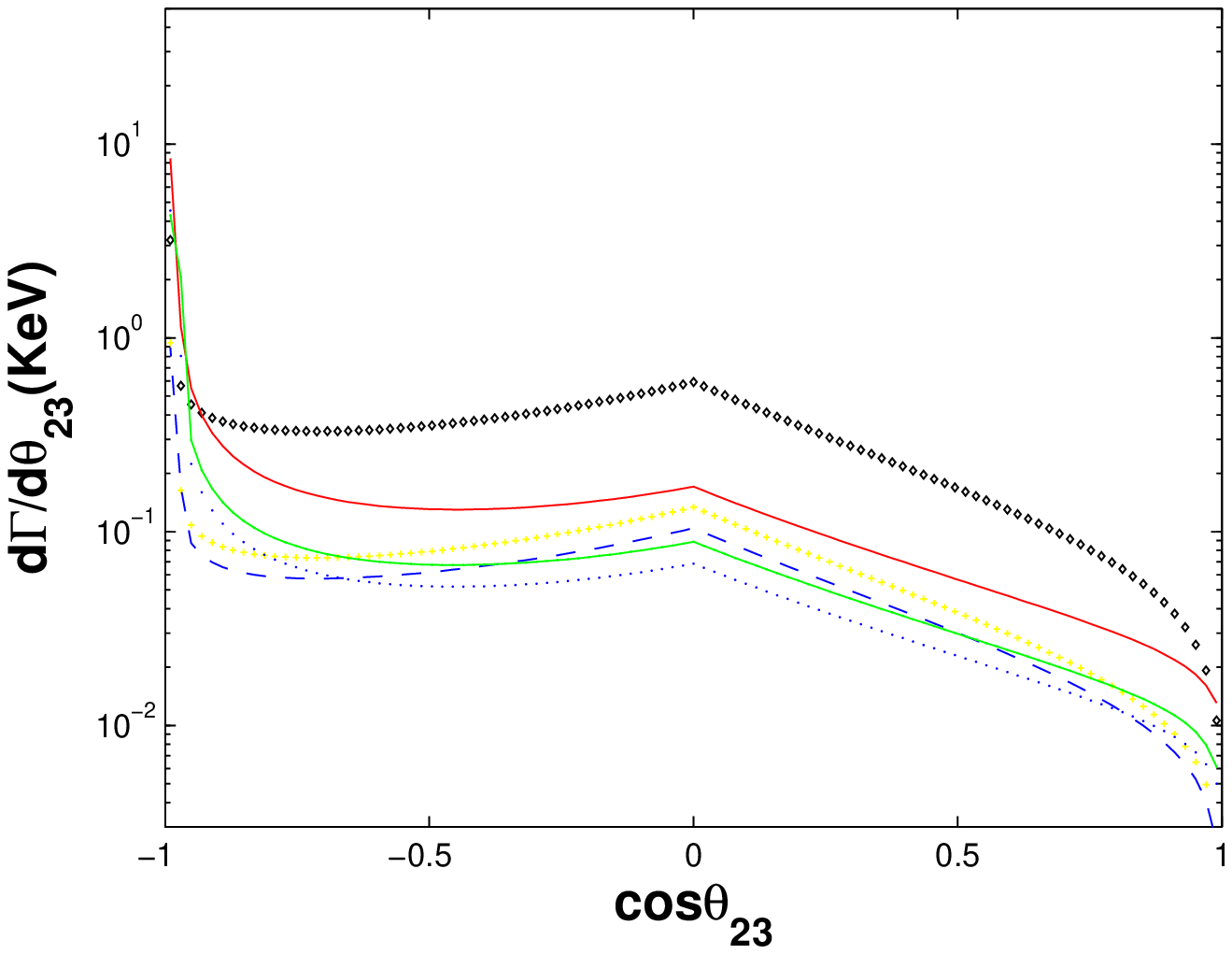}
\caption{(color online). Differential decay widths $d\Gamma/ds_1$, $d\Gamma/ds_2$, $d\Gamma/dcos\theta_{12}$ and $d\Gamma/dcos\theta_{23}$ for $ H^{0}\rightarrow |(b\bar{c})[n]\rangle +c\bar{b}$, where the diamond line, the cross line, the dashed line, the solid line, the dotted line and the dash-dotted line are for $|(b\bar{c})[1S]\rangle$, $|(b\bar{c})[2S]\rangle$, $|(b\bar{c})[3S]\rangle$, $|(b\bar{c})[1P]\rangle$, $|(b\bar{c})[2P]\rangle$ and $|(b\bar{c})[3P]\rangle$, respectively.} \label{H(bc)dsdcos}
\end{figure}
\begin{figure}[htbp]
\includegraphics[width=0.23\textwidth]{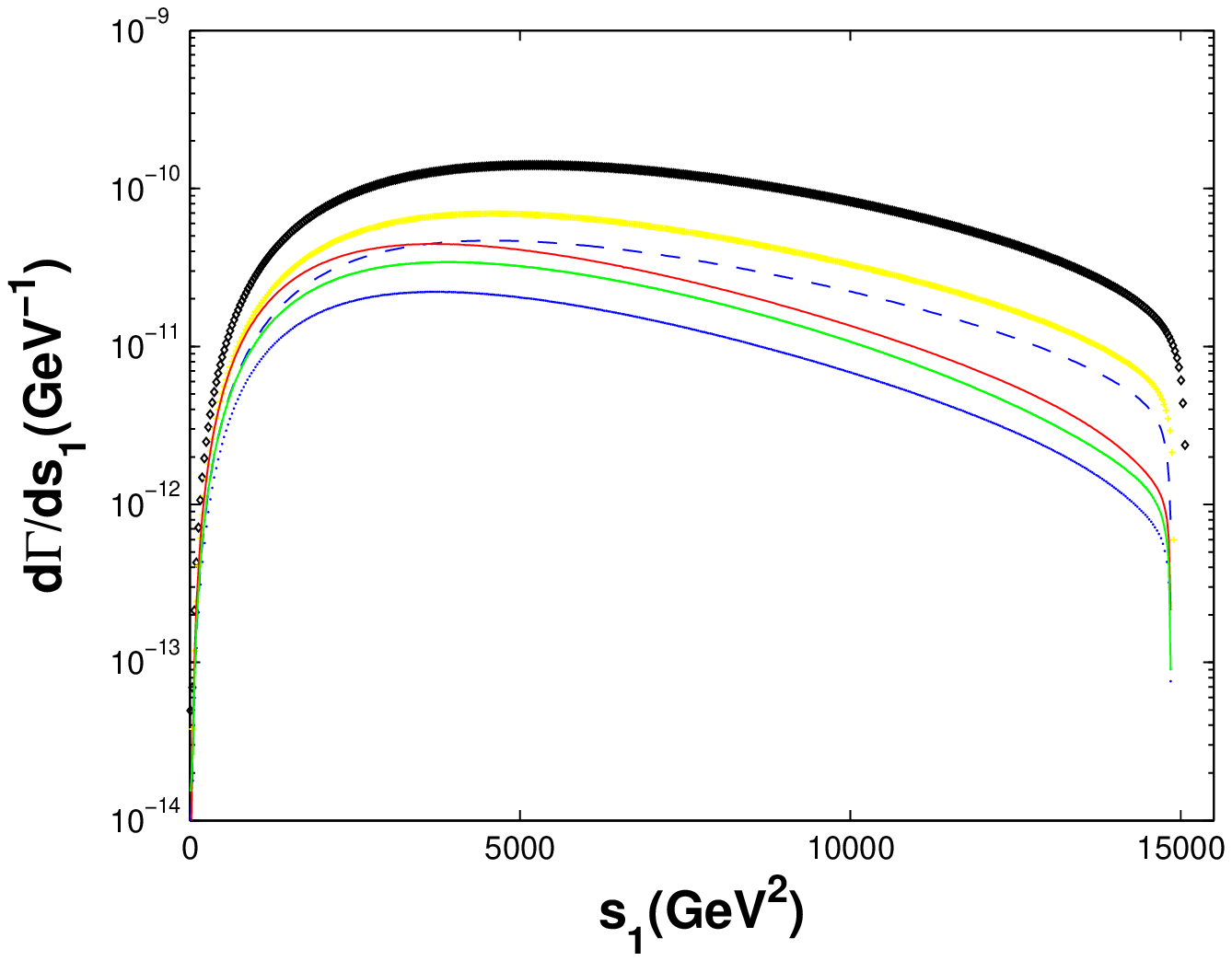}
\includegraphics[width=0.23\textwidth]{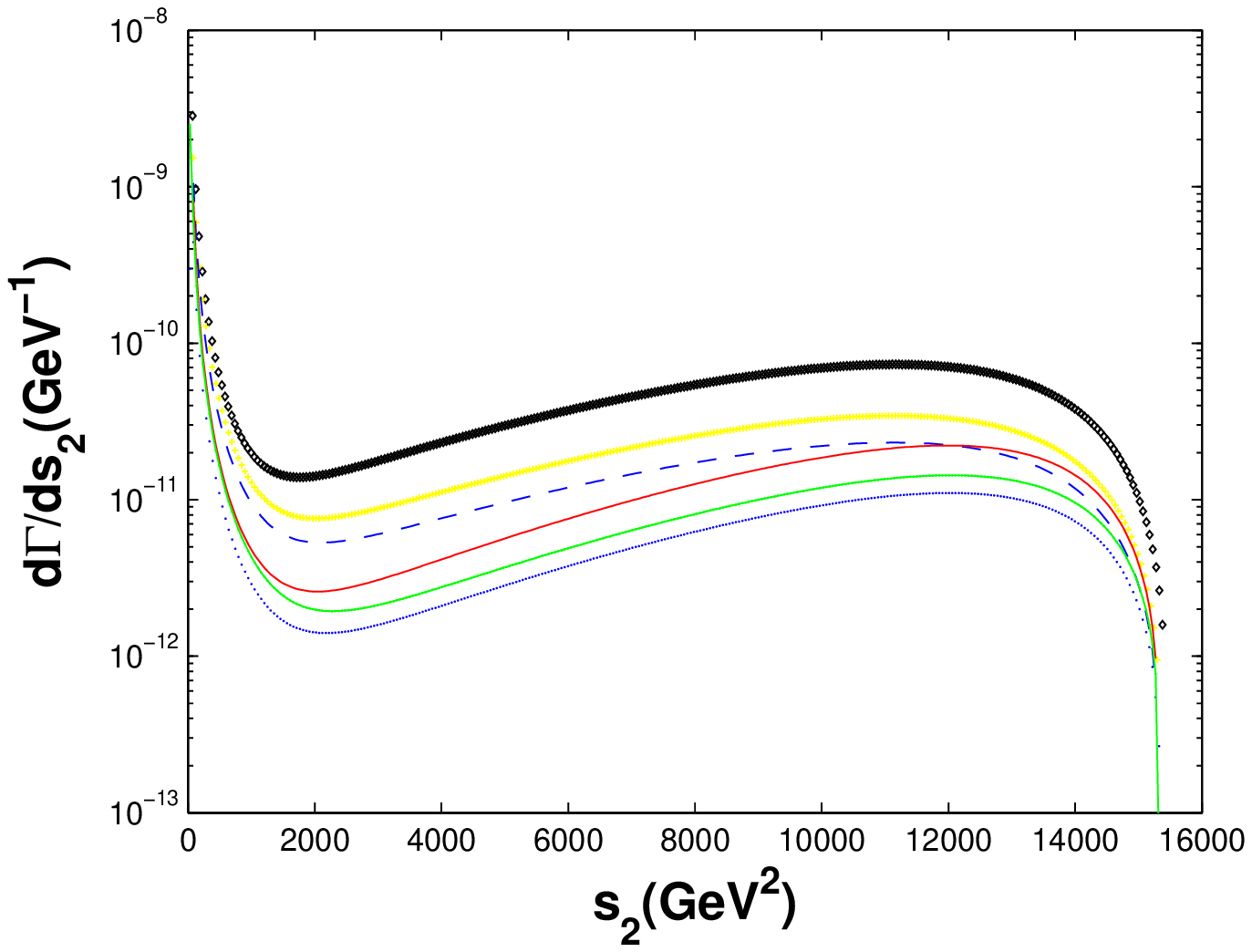}
\includegraphics[width=0.23\textwidth]{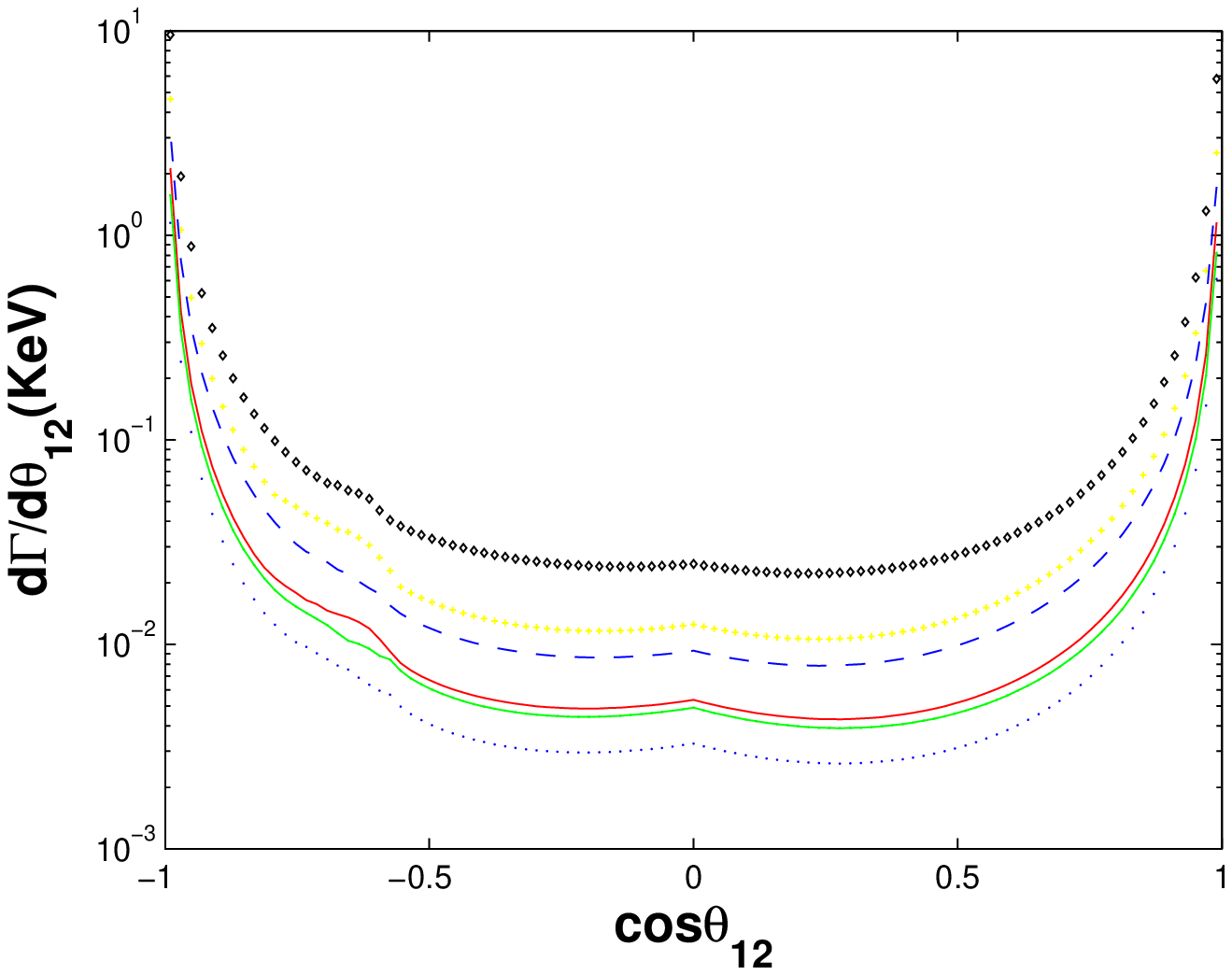}
\includegraphics[width=0.23\textwidth]{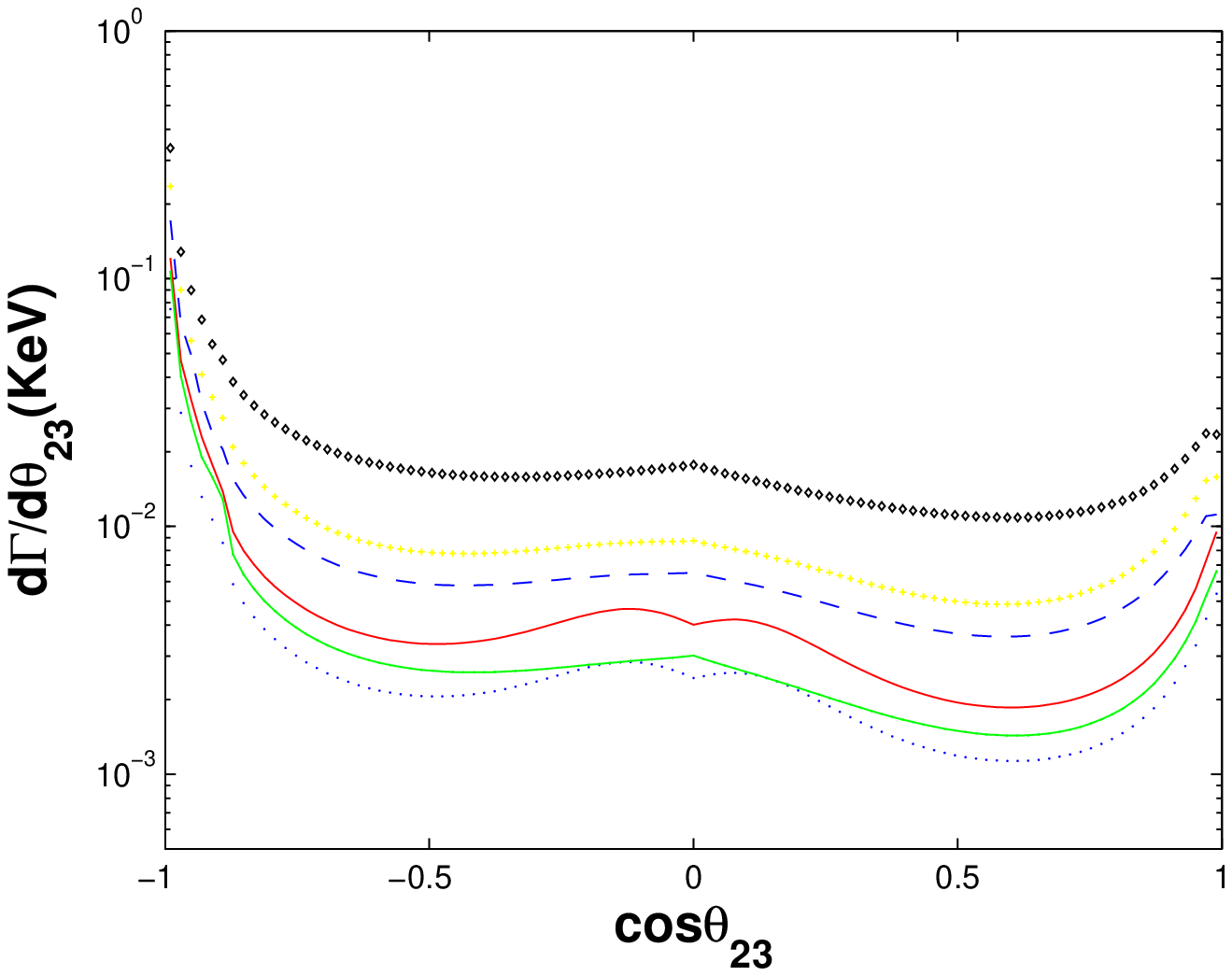}
\caption{(color online). Differential decay widths $d\Gamma/ds_1$, $d\Gamma/ds_2$, $d\Gamma/dcos\theta_{12}$ and $d\Gamma/dcos\theta_{23}$ for $ H^{0}\rightarrow |(c\bar{c})[n]\rangle +c\bar{c}$, where the diamond line, the cross line, the dashed line, the solid line, the dotted line and the dash-dotted line are for $|(c\bar{c})[1S]\rangle$, $|(c\bar{c})[2S]\rangle$, $|(c\bar{c})[3S]\rangle$, $|(c\bar{c})[1P]\rangle$, $|(c\bar{c})[2P]\rangle$ and $|(c\bar{c})[3P]\rangle$, respectively.} \label{H(cc)dsdcos}
\end{figure}
\begin{figure}[htbp]
\includegraphics[width=0.23\textwidth]{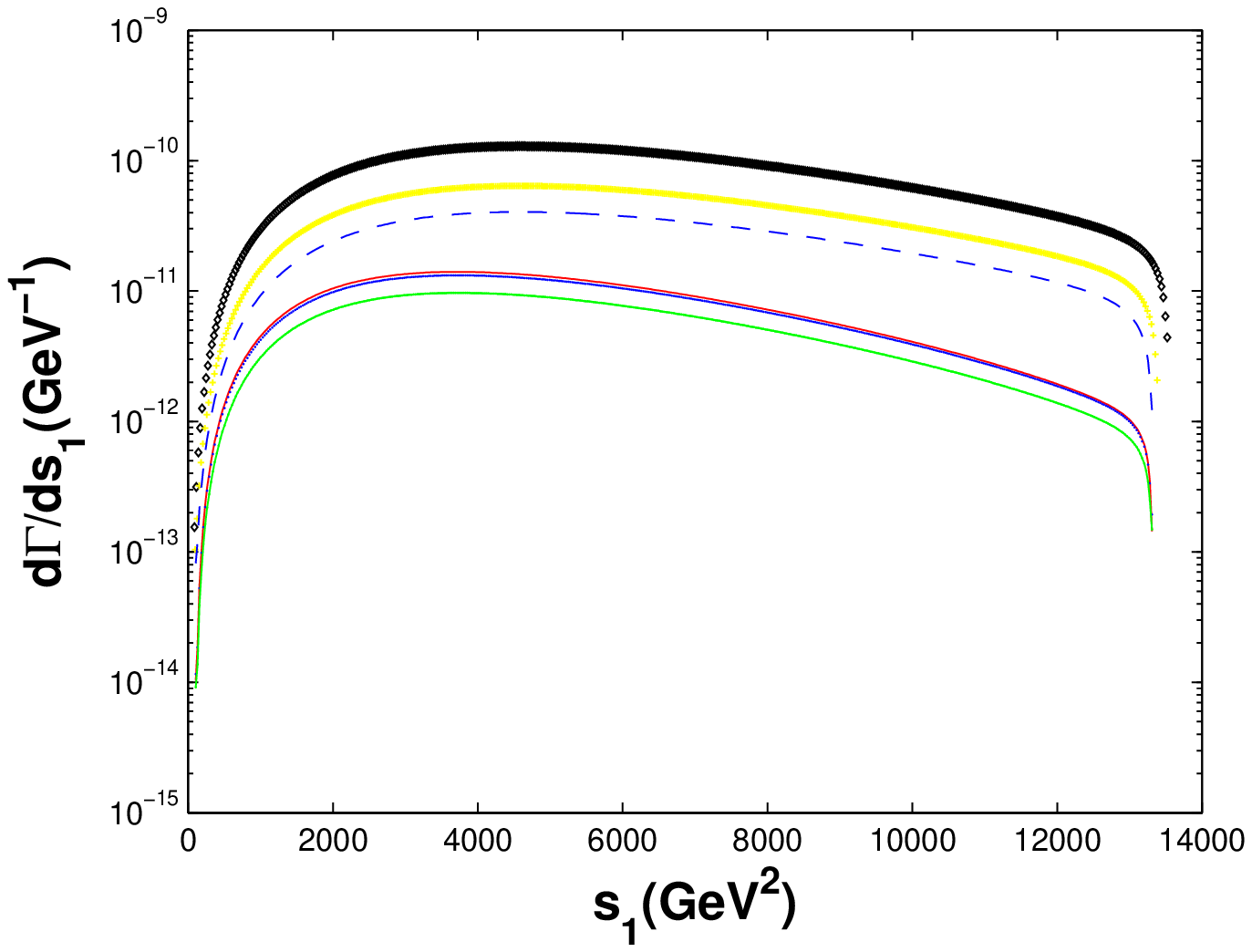}
\includegraphics[width=0.23\textwidth]{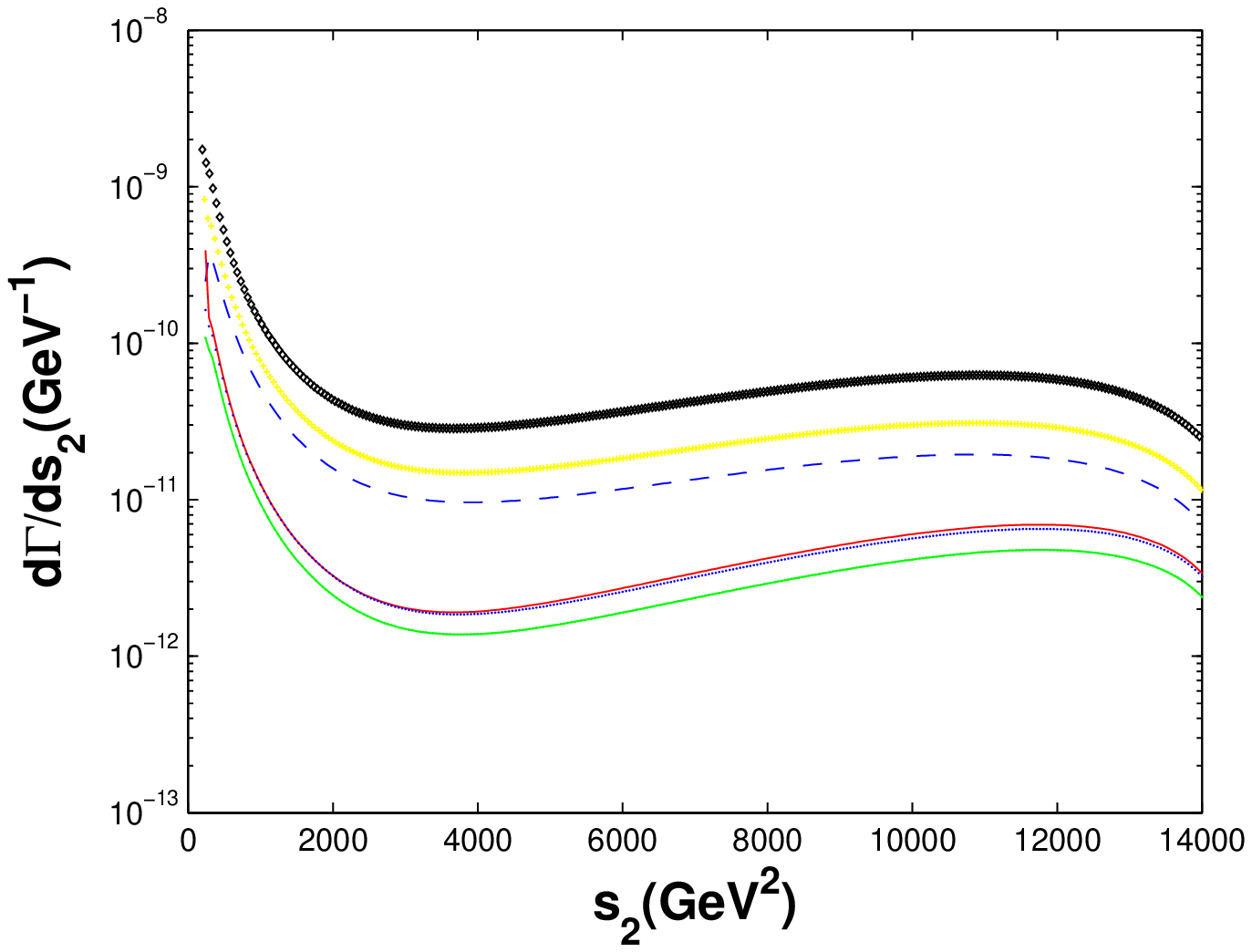}
\includegraphics[width=0.23\textwidth]{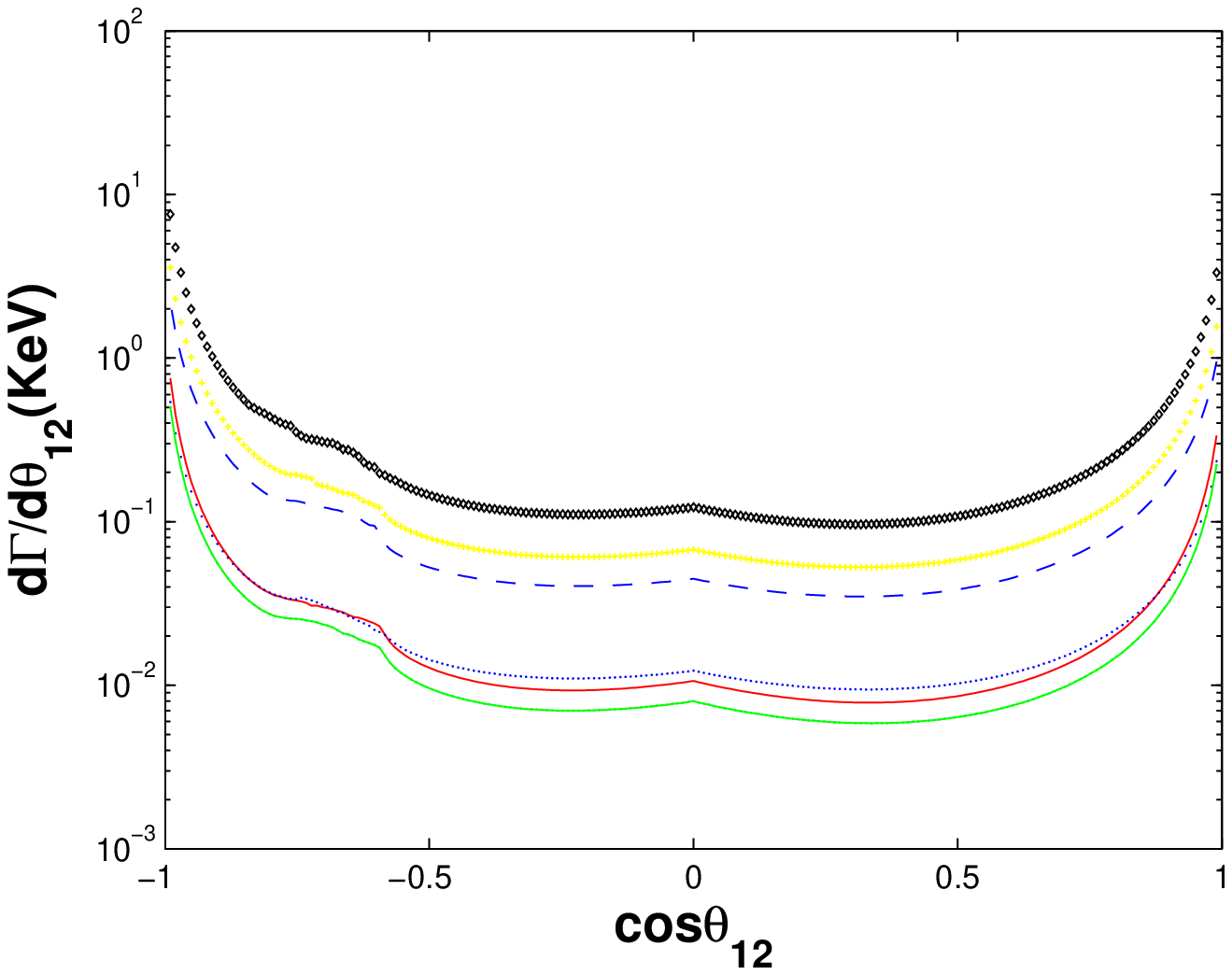}
\includegraphics[width=0.23\textwidth]{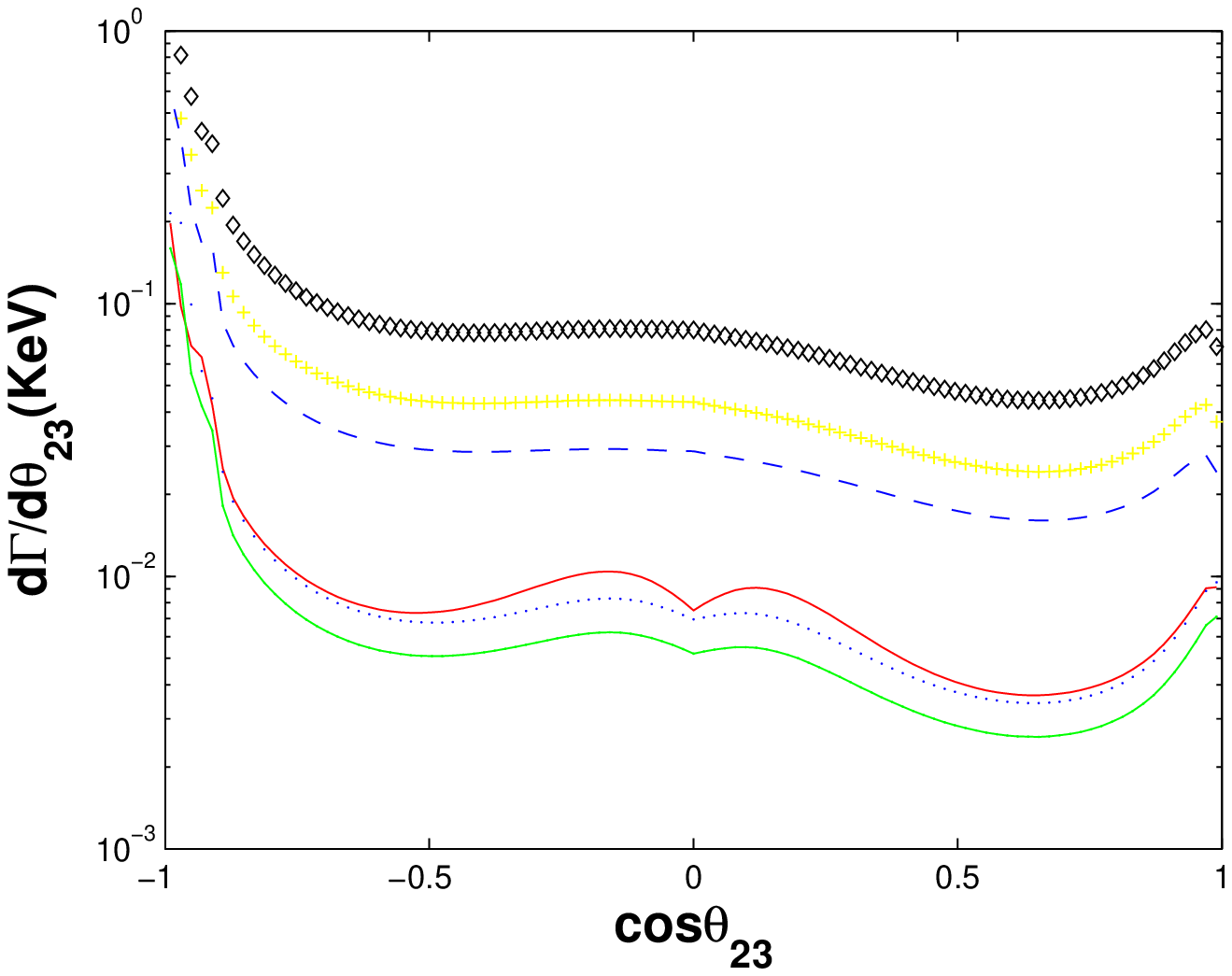}
\caption{(color online). Differential decay widths $d\Gamma/ds_1$, $d\Gamma/ds_2$, $d\Gamma/dcos\theta_{12}$ and $d\Gamma/dcos\theta_{23}$ for $ H^{0}\rightarrow |(b\bar{b})[n]\rangle +b\bar{b}$, where the diamond line, the cross line, the dashed line, the solid line, the dotted line and the dash-dotted line are for $|(b\bar{b})[1S]\rangle$, $|(b\bar{c})[2S]\rangle$, $|(b\bar{b})[3S]\rangle$, $|(b\bar{b})[1P]\rangle$, $|(b\bar{b})[2P]\rangle$ and $|(b\bar{b})[3P]\rangle$, respectively.} \label{H(bb)dsdcos}
\end{figure}

The decay widths for the production of high excited quarkonia through $H^0$ semi-exclusive decays, i.e., $H^0\rightarrow |(b\bar{c})[n]\rangle+\bar{b}c$, $H^0\rightarrow |(c\bar{c})[n]\rangle+\bar{c}c$ and $H^0\rightarrow |(b\bar{b})[n]\rangle+\bar{b}b$, are listed in Tables \ref{tabrpa}, \ref{tabrpb} and \ref{tabrpc}. Here we adopt the BT-potential model for the non-perturbative hadronic matrix elements. If the input parameters of the Ref. \cite{jjcq} and  \cite{lx1} are adopted, values are consistent with the leading-order results for $1S$ and $1P$-wave states of those papers.

From Tables~\ref{tabrpa}$\sim$\ref{tabrpc}, it is shown that, in addition to the ground $1S$-level states, the high excited states of $nS$ and $nP$-wave states of $|(Q\bar{Q'})[n]\rangle$ quarkonia  can provide sizable contributions to the total decay widths. Here $nS$ ($n=1, 2, 3,4$) stands for the summed decay widths of $n^1S_0$ and $n^3S_1$ at the same $n$th level, and $nP$ stands for the summed decay width of $n^1P_1$ and $n^3P_J$ ($J=0,1,2$) at the same $n$th level.
\begin{itemize}
\item For $|(b\bar{c})[n]\rangle$ quarkonium production in $H^0$ boson semi-exclusive decays, the decay widths for $2S$, $3S$, $4S$, $1P$, $2P$, $3P$ and $4P$-wave states are about $18.2\%$, $13.1\%$, $12.0\%$, $15.6\%$, $5.50\%$, $6.51\%$ and $6.40\%$ of the decay width of the $|(b\bar{c})[1S]\rangle$ quarkonium production, respectively.
\end{itemize}
\begin{itemize}
\item For  charmonium production in $H^0$ semi-exclusive decays, the decay widths for $2S$, $3S$, $4S$, $1P$, $2P$, $3P$ and $4P$-wave states are about $47.0\%$, $31.7\%$, $24.7\%$, $23.7\%$, $11.7\%$, $15.1\%$, and $15.6\%$ of the decay width of the $|(c\bar{c})[1S]\rangle$ quarkonium production, respectively.
\end{itemize}
\begin{itemize}
\item For bottomonium production in $H^0$ semi-exclusive decays, the decay widths for $2S$, $3S$, $4S$, $1P$, $2P$, $3P$ and $4P$-wave states are about $49.5\%$, $31.2\%$, $18.7\%$, $10.0\%$, $8.34\%$, $5.97\%$ and $3.85\%$ of the decay width of the $|(b\bar{b})[1S]\rangle$ quarkonium production, respectively.
\end{itemize}
To further compare the contributions of the ground and high excited $|(b\bar{c})[n]\rangle$, $|(c\bar{c})[n]\rangle$ and $|(b\bar{b})[n]\rangle$ states, we present the differential distributions $d\Gamma/ds_{1}$, $d\Gamma/ds_{2}$, $d\Gamma/dcos\theta_{12}$, and $d\Gamma/dcos\theta_{23}$ for the $H^0\to |(b\bar{c})[n]\rangle +\bar{b}c$,  $H^0\to |(c\bar{c})[n]\rangle +\bar{c}c$ and  $H^0\to |(b\bar{b})[n]\rangle +\bar{b}b$ processes in Figs. \ref{H(bc)dsdcos}$\sim$\ref{H(bb)dsdcos}.  Here, $s_{1}=(q_1+q_2)^2$, $s_{2}=(q_1+q_3)^2$, $\theta_{12}$ is the angle between $\vec{q}_1$ and $\vec{q}_2$, and $\theta_{23}$ is that between $\vec{q}_2$ and $\vec{q}_3$.  Again, these figures show explicitly that, in almost the entire kinematical region, the high excited Fock states can provide sizable contributions in comparison with the lower Fock state $|(b\bar{c})[1S]\rangle$.
In general, the line shapes of the same distribution are similar for the three channels. And comparatively, the curves of charmonium and bottomonium are flatter than those of $|(b\bar{c})[n]\rangle$ quarkonium.

\begin{figure*}[htb]
\includegraphics[width=0.23\textwidth]{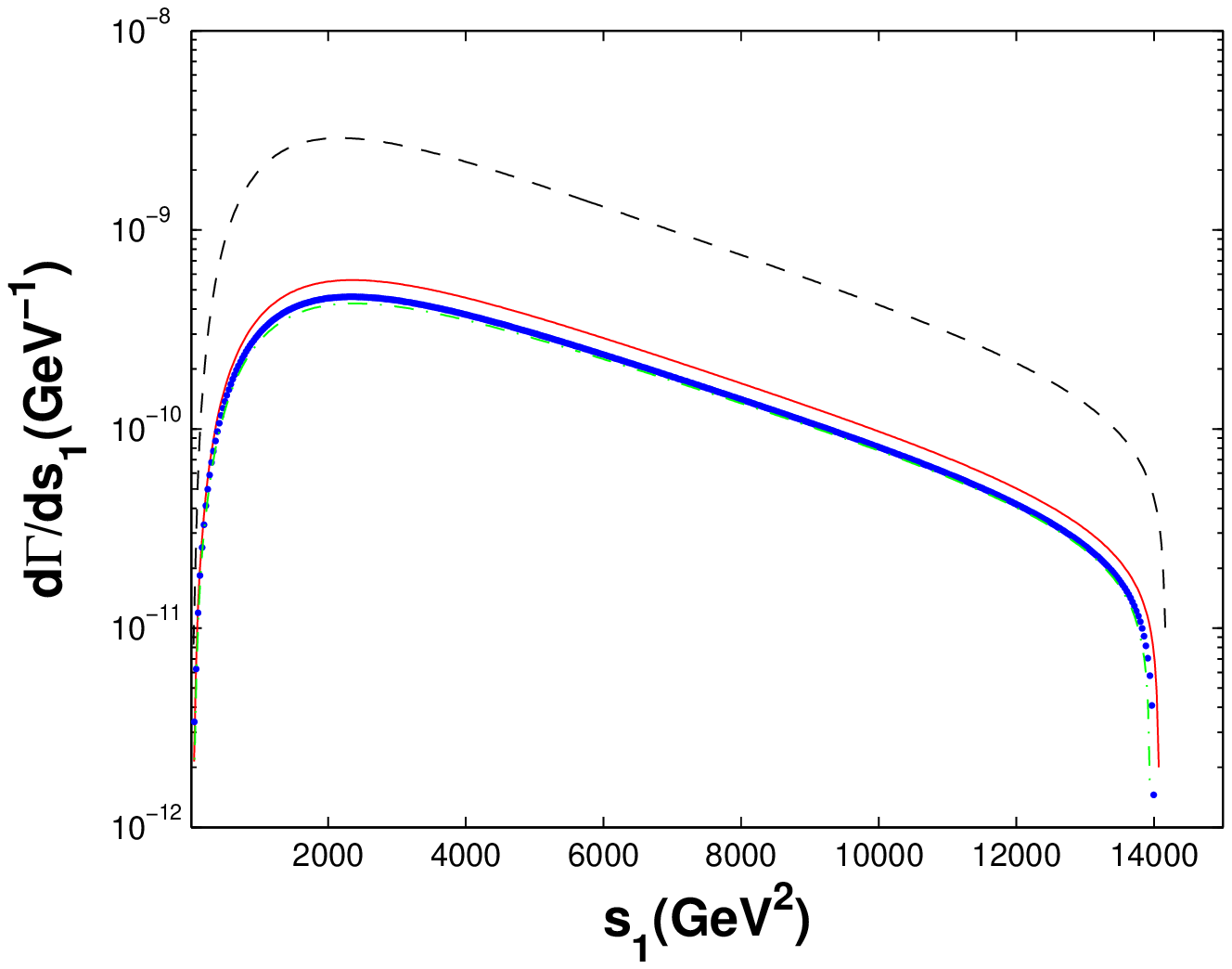}
\includegraphics[width=0.23\textwidth]{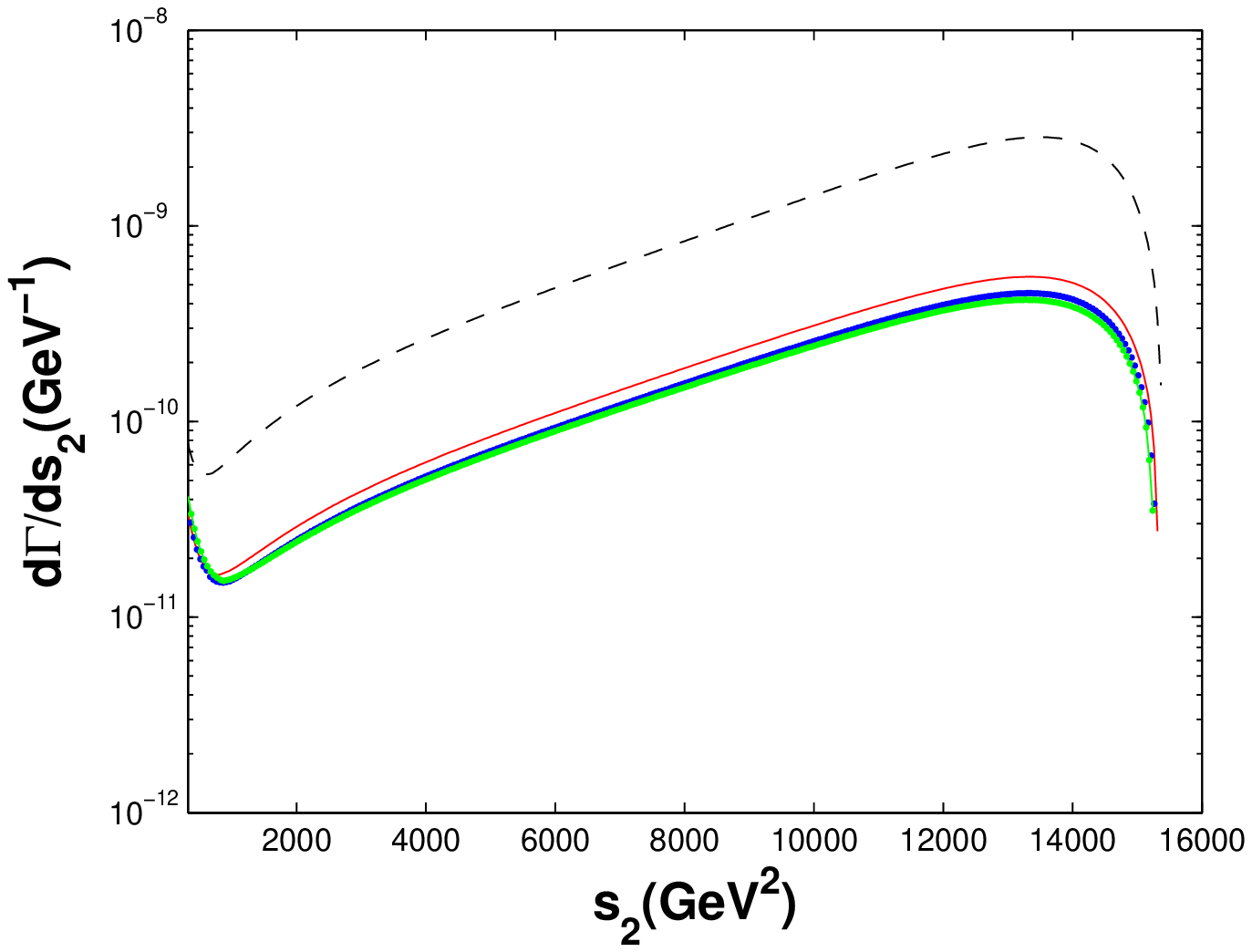}
\includegraphics[width=0.23\textwidth]{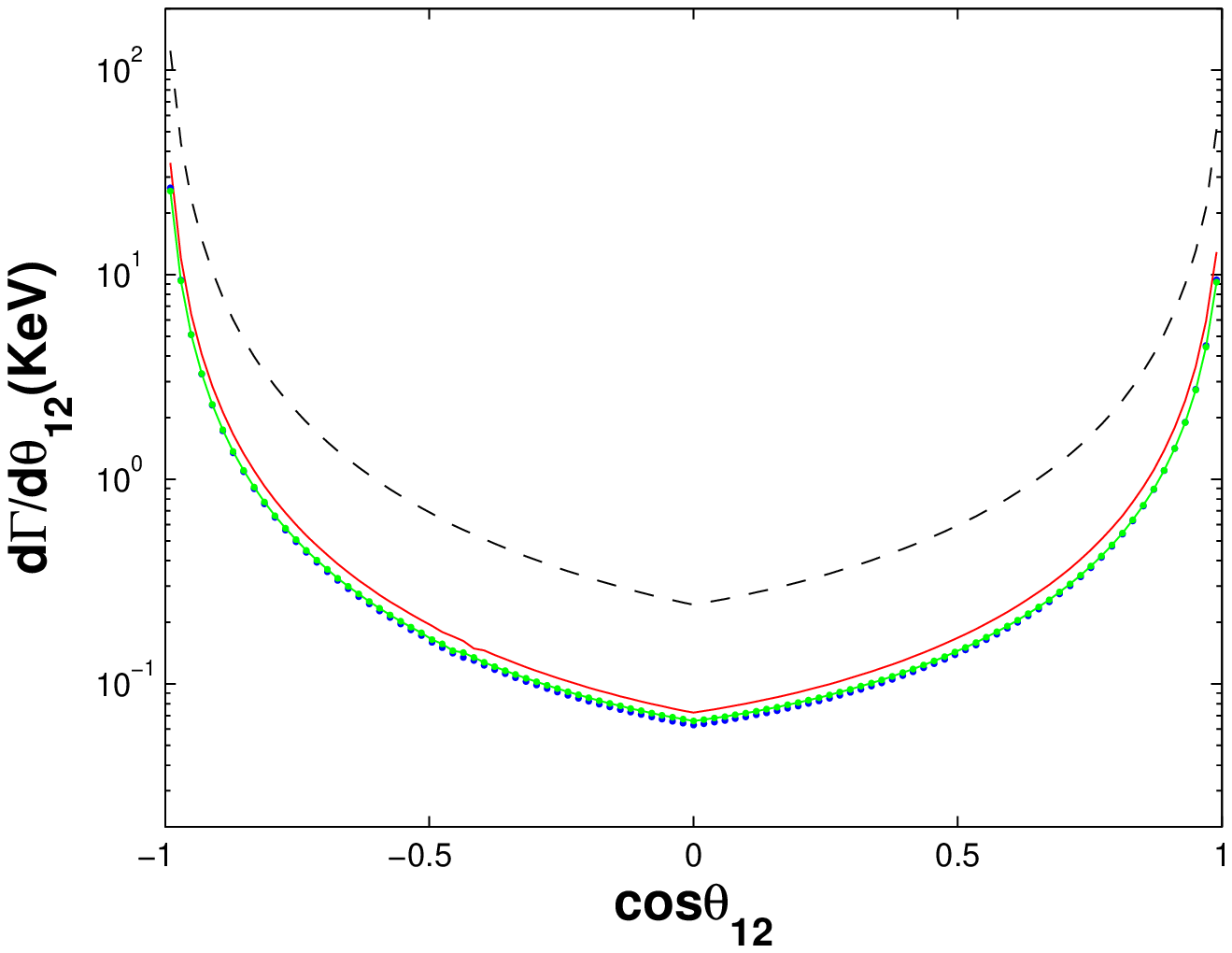}
\includegraphics[width=0.23\textwidth]{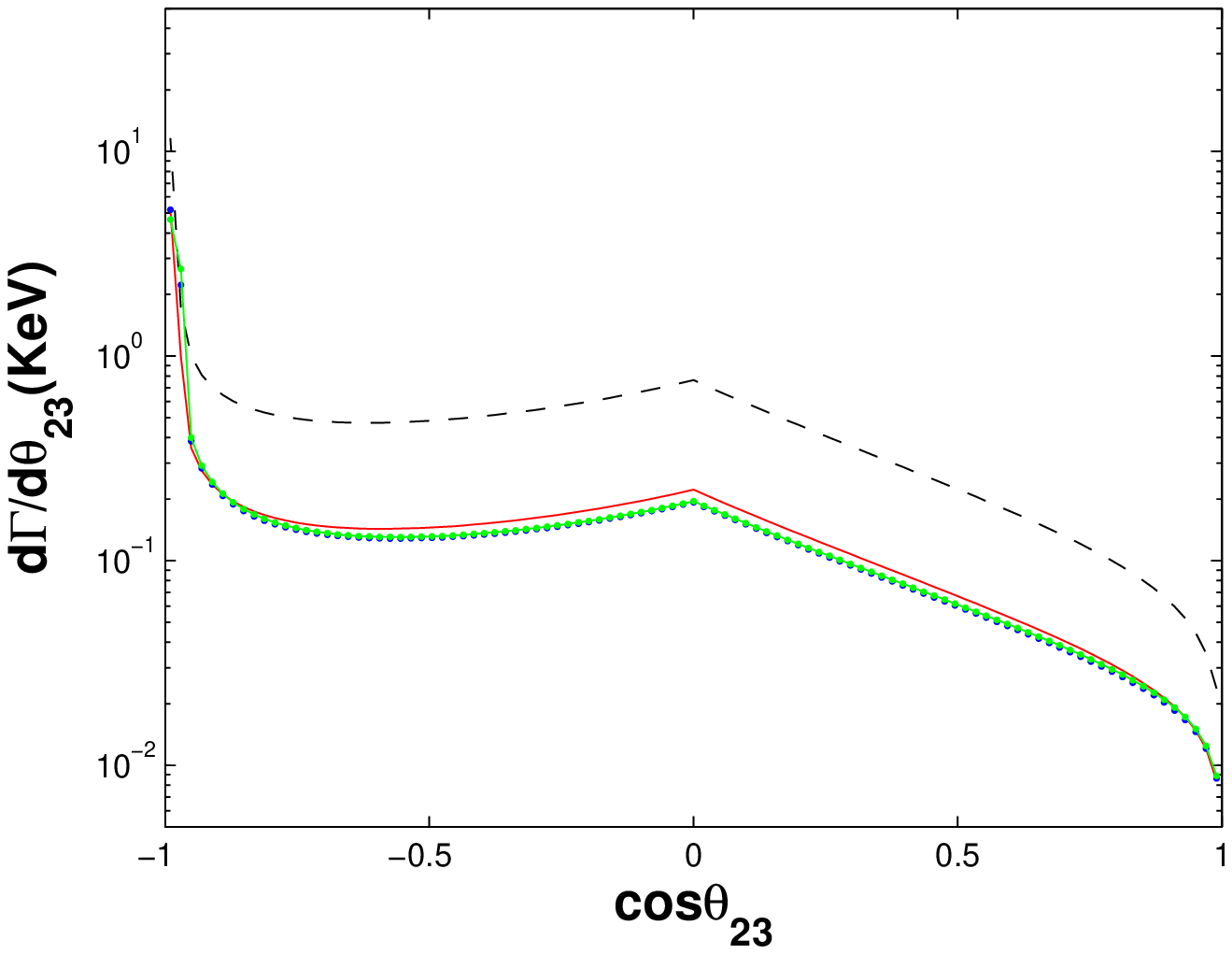}
\caption{(color online). Differential decay widths $d\Gamma/ds_1$, $d\Gamma/ds_2$, $d\Gamma/dcos\theta_{12}$ and $d\Gamma/dcos\theta_{23}$ for $ H^{0}\rightarrow |(b\bar{c})[n]\rangle +c\bar{b}$, where the dashed line, the solid line, the dotted line and the dash-dotted line are for $|(b\bar{c})[1]\rangle$, $|(b\bar{c})[2]\rangle$, $|(b\bar{c})[3]\rangle$ and $|(b\bar{c})[4]\rangle$, respectively.} \label{H(bc)dsdcosSUM}
\end{figure*}
\begin{figure*}[htb]
\includegraphics[width=0.23\textwidth]{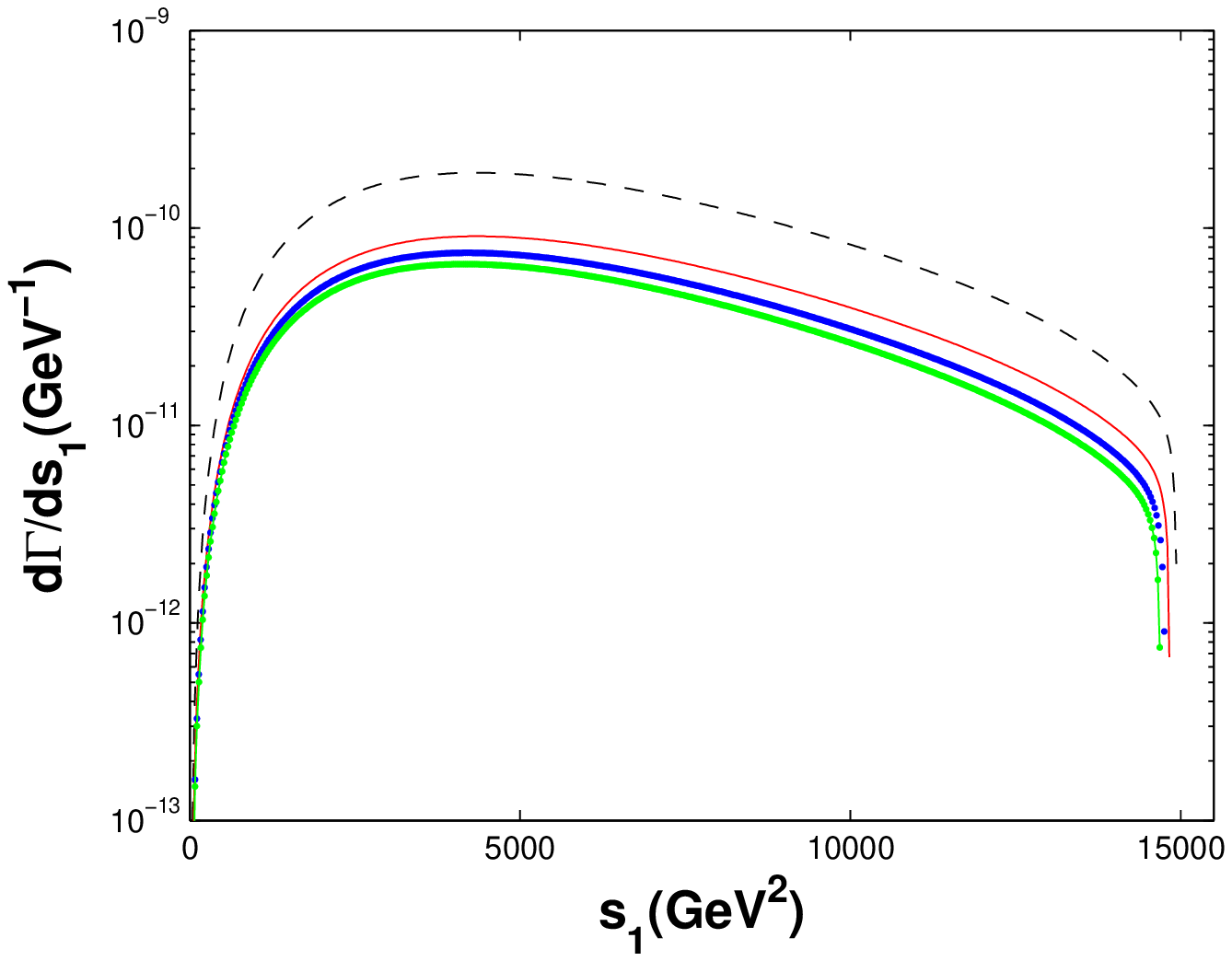}
\includegraphics[width=0.23\textwidth]{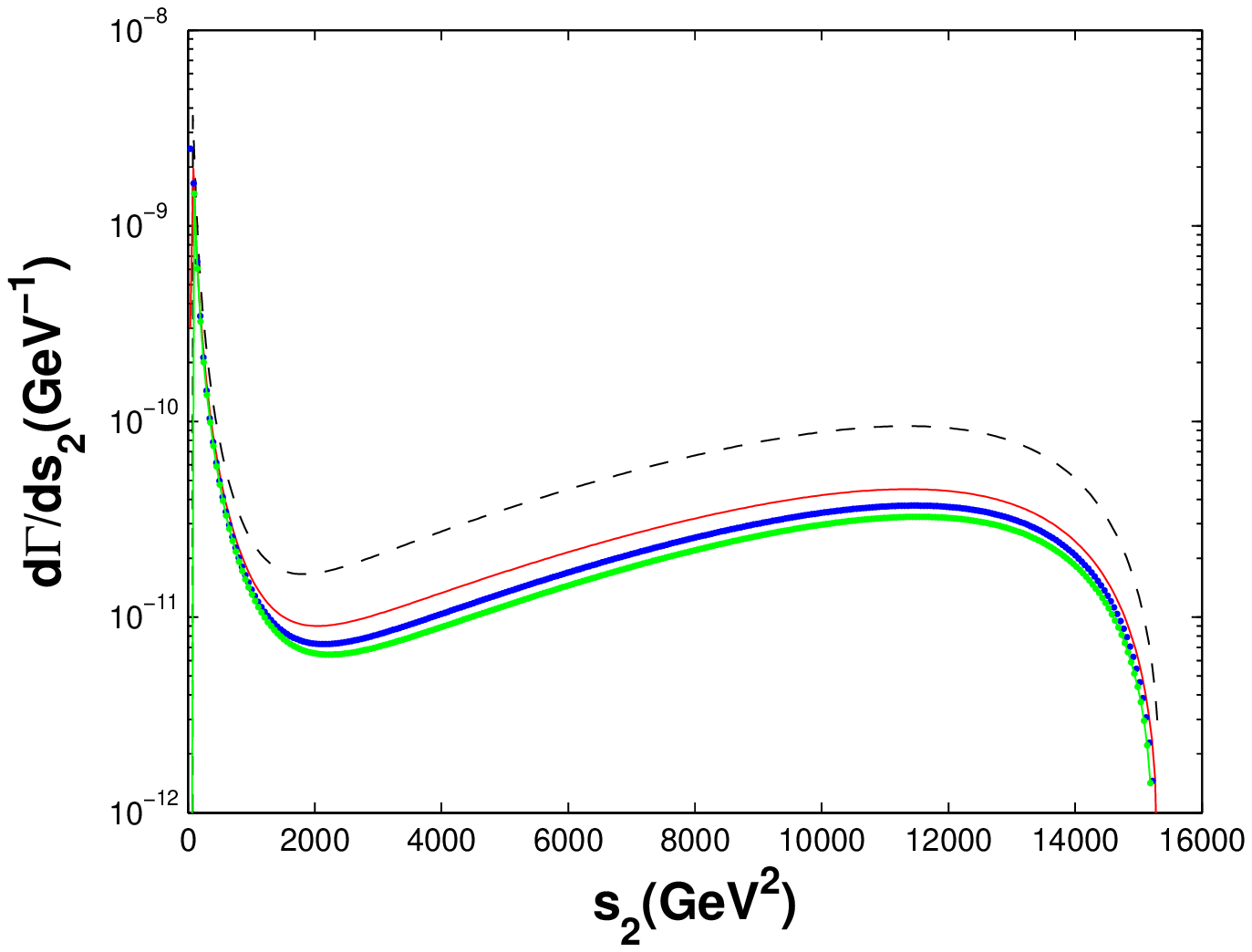}
\includegraphics[width=0.23\textwidth]{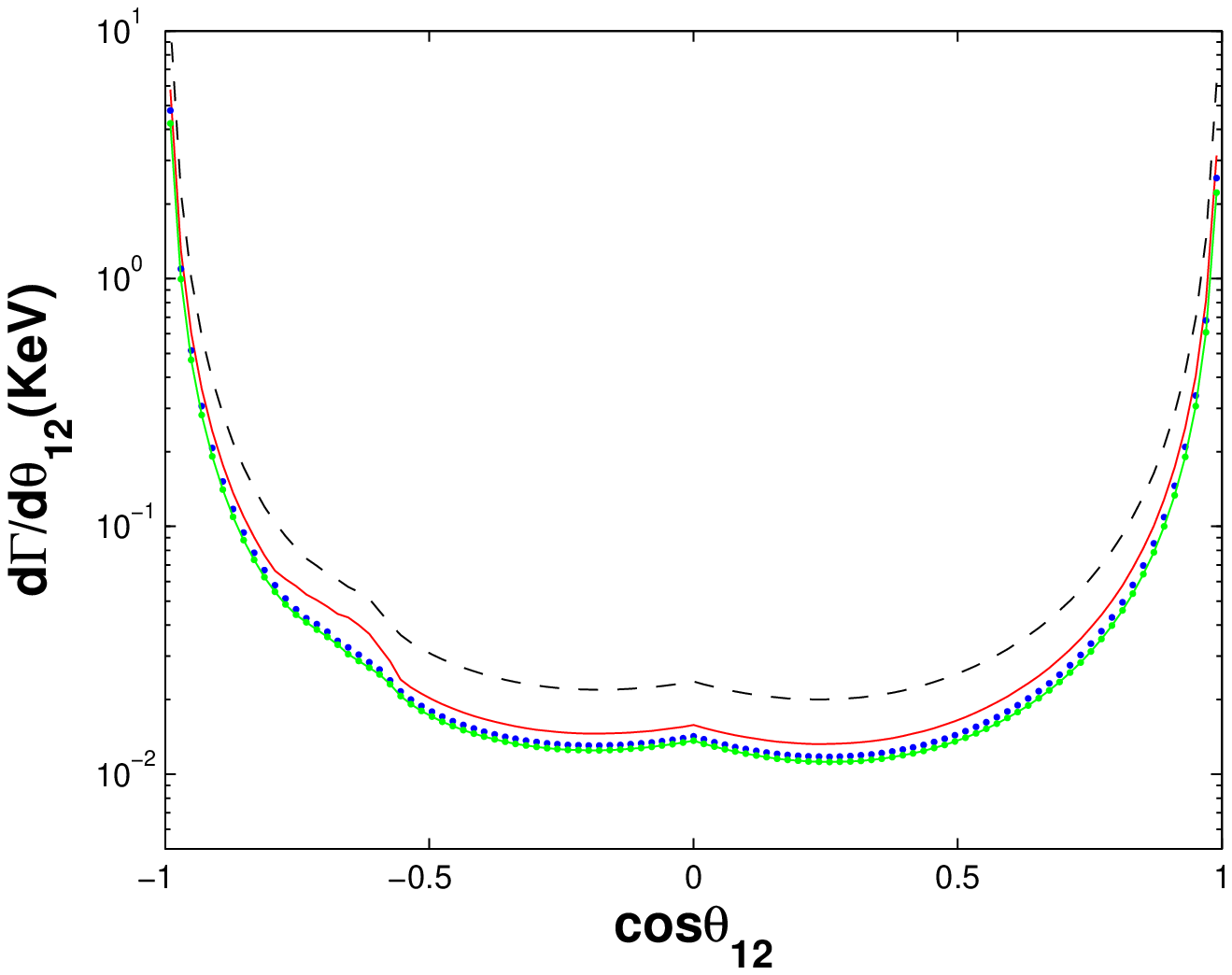}
\includegraphics[width=0.23\textwidth]{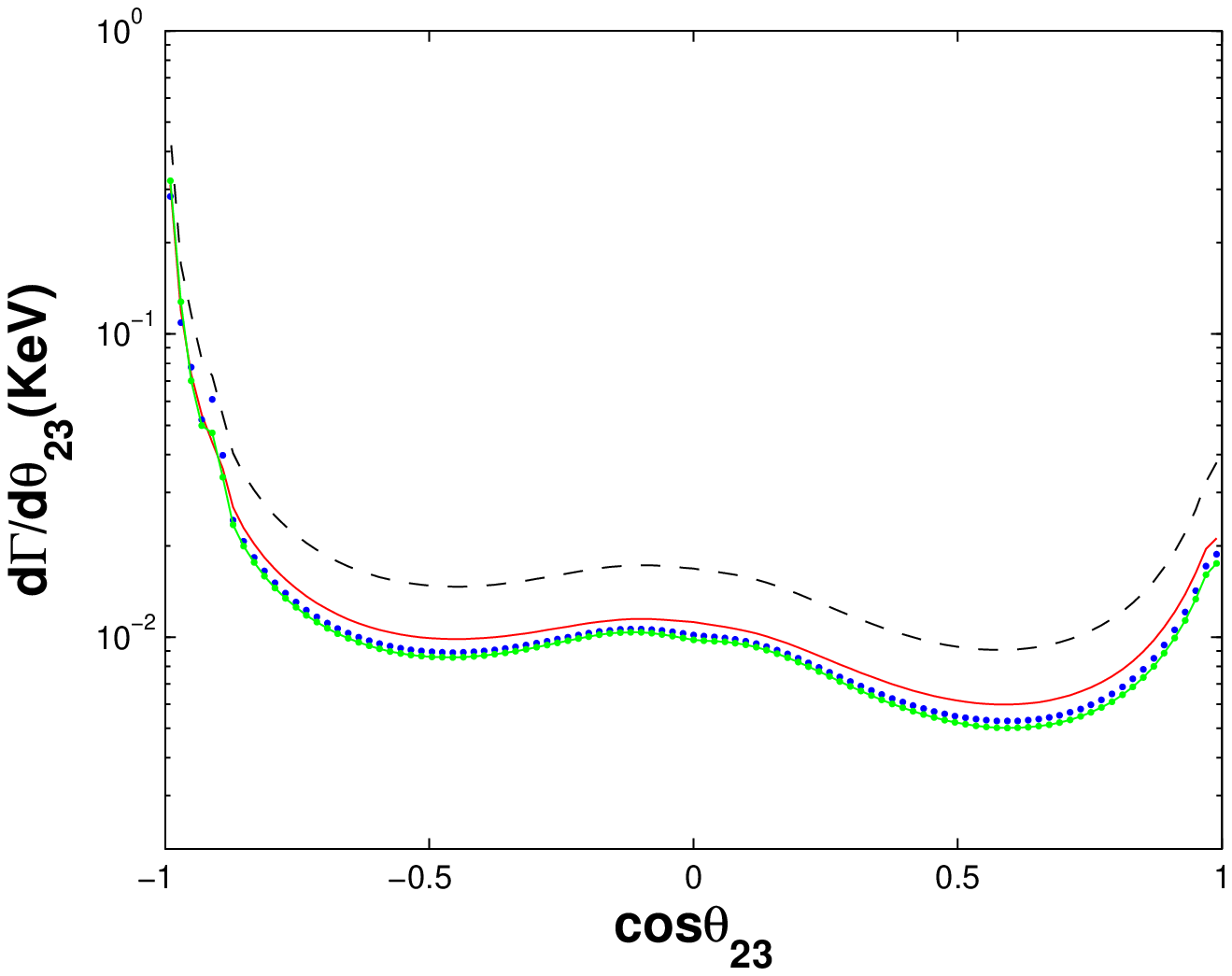}
\caption{(color online). Differential decay widths $d\Gamma/ds_1$, $d\Gamma/ds_2$, $d\Gamma/dcos\theta_{12}$ and $d\Gamma/dcos\theta_{23}$ for $ H^{0}\rightarrow |(c\bar{c})[n]\rangle +c\bar{c}$, where the dashed line, the solid line, the dotted line and the dash-dotted line are for $|(c\bar{c})[1]\rangle$, $|(c\bar{c})[2]\rangle$, $|(c\bar{c})[3]\rangle$ and $|(c\bar{c})[4]\rangle$, respectively.} \label{H(cc)dsdcosSUM}
\end{figure*}
\begin{figure*}[htb]
\includegraphics[width=0.23\textwidth]{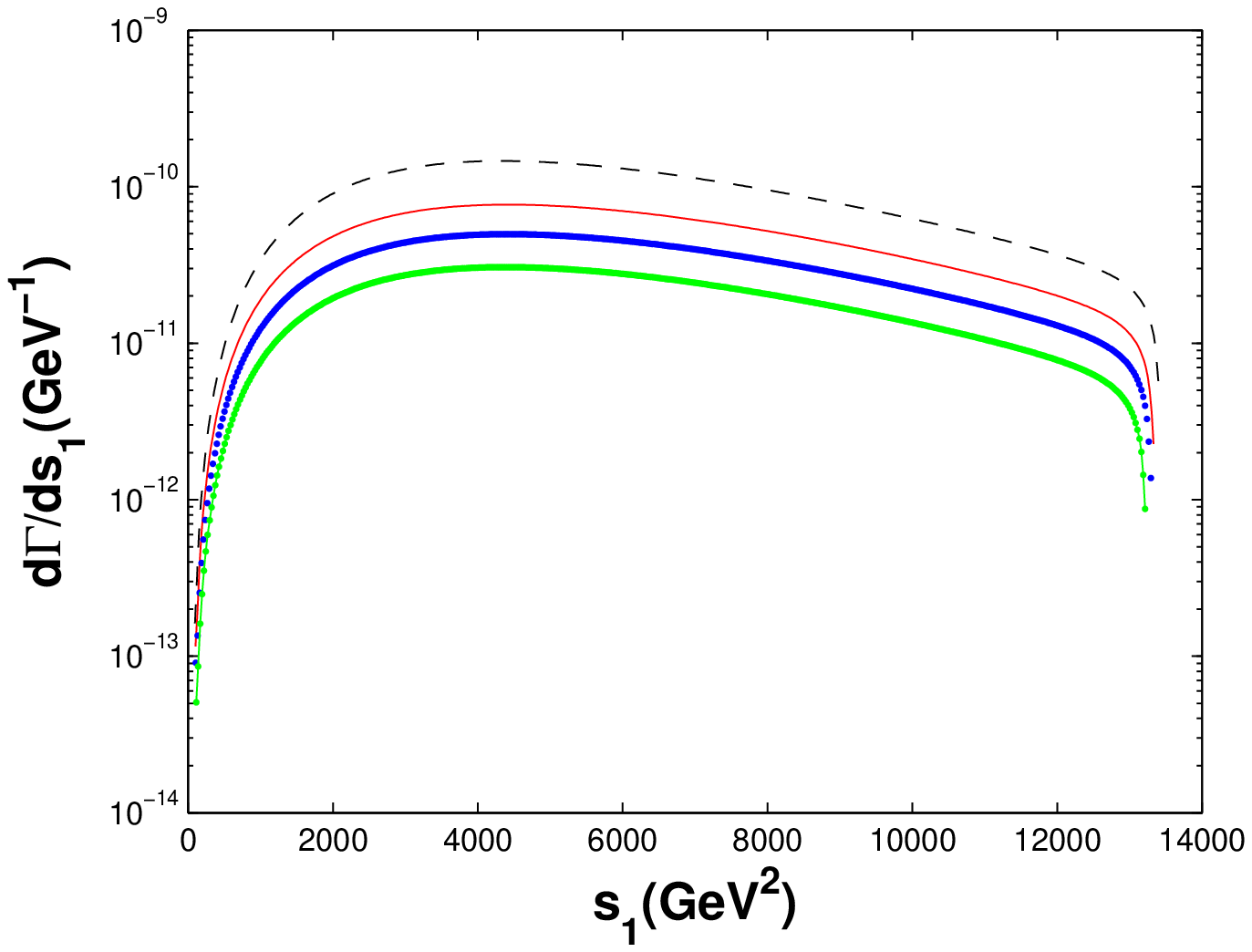}
\includegraphics[width=0.23\textwidth]{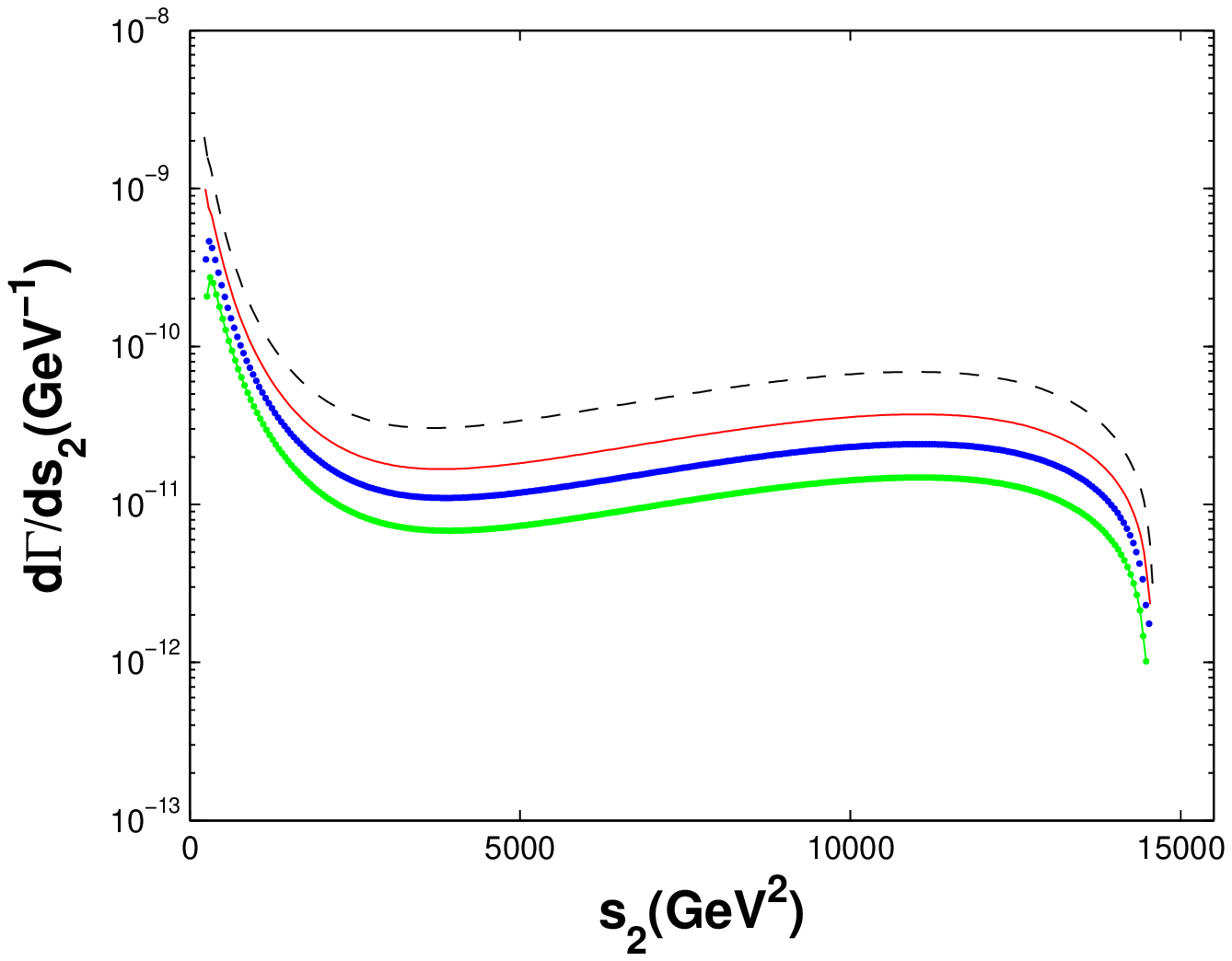}
\includegraphics[width=0.23\textwidth]{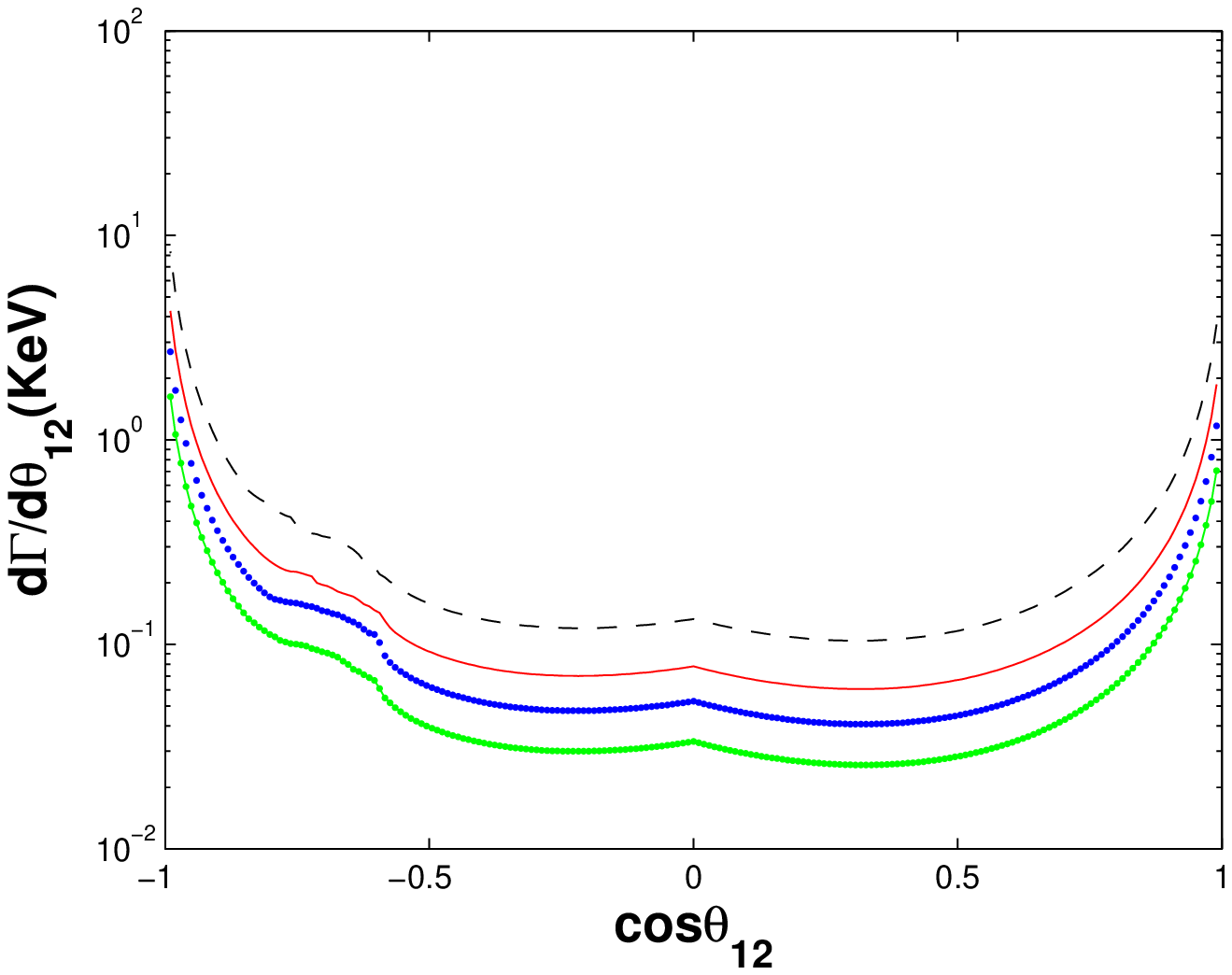}
\includegraphics[width=0.23\textwidth]{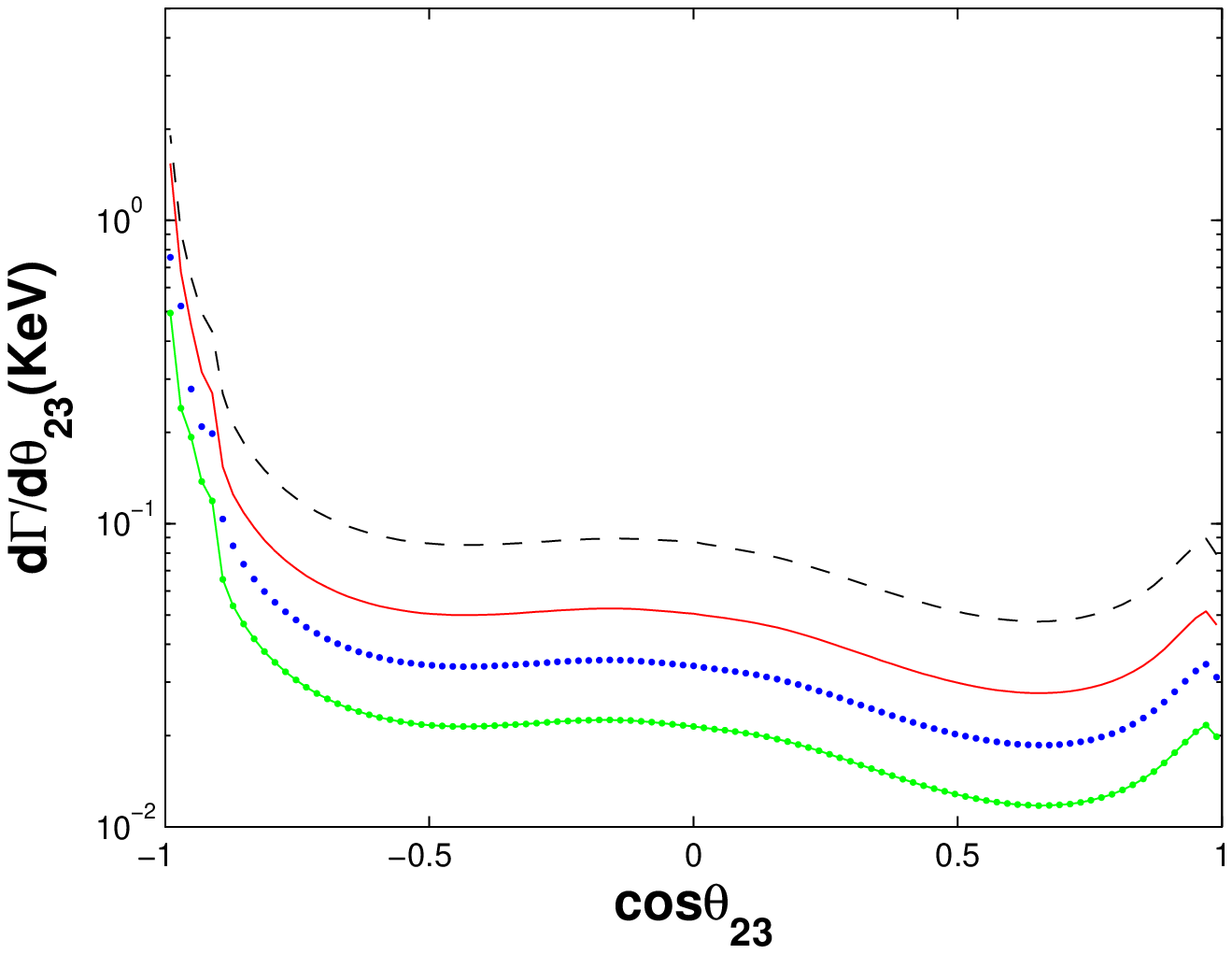}
\caption{(color online). Differential decay widths $d\Gamma/ds_1$, $d\Gamma/ds_2$, $d\Gamma/dcos\theta_{12}$ and $d\Gamma/dcos\theta_{23}$ for $ H^{0}\rightarrow |(b\bar{b})[n]\rangle +b\bar{b}$, where the dashed line, the solid line, the dotted line and the dash-dotted line are for $|(b\bar{b})[1]\rangle$, $|(b\bar{b})[2]\rangle$, $|(b\bar{b})[3]\rangle$ and $|(b\bar{b})[4]\rangle$, respectively.} \label{H(bb)dsdcosSUM}
\end{figure*}

Alternatively,  using $[n]$ to represent the sum of decay widths of $n^1S_0$, $n^3S_1$, $n^1P_1$ and $n^3P_J$ ($J=0,1,2$) at the same $n$th level, one also find that the high excited Fock states make significant contributions.\begin{itemize}
\item For $|(b\bar{c})[n]\rangle$ quarkonium production in $H^0$ boson decays, the decay widths for $|(b\bar{c})[2]\rangle$, $|(b\bar{c})[3]\rangle$ and $|(b\bar{c})[4]\rangle$ states are about $22.3\%$, $18.4\%$, $17.2\%$ of the decay width of the $|(b\bar{c})[1]\rangle$ quarkonium production, respectively.
\end{itemize}
\begin{itemize}
\item For  charmonium production in $H^0$ boson decays, the decay widths for $|(c\bar{c})[2]\rangle$, $|(c\bar{c})[3]\rangle$ and $|(c\bar{c})[4]\rangle$ states are about $47.5\%$, $35.6\%$, $32.5\%$ of the decay width of the $|(c\bar{c})[1]\rangle$ quarkonium production, respectively.
\end{itemize}
\begin{itemize}
\item For bottomonium production in $H^0$ boson decays, the decay widths for $|(b\bar{b})[2]\rangle$, $|(b\bar{b})[3]\rangle$ and $|(b\bar{b})[4]\rangle$ states are about $52.6\%$, $33.7\%$, $20.5\%$ of the decay width of the $|(b\bar{b})[1]\rangle$ quarkonium production, respectively.
\end{itemize}
 It is found that, the decay widths of $|(b\bar{c})[n]\rangle$ meson are the largest among the three channels, yet the proportions of the high excited states are much smaller than those for charmonium and bottomonium. In  Figs. \ref{H(bc)dsdcosSUM}$\sim$\ref{H(bb)dsdcosSUM}, we also display the differential distributions by summing up the decay widths of various Fock states at the same $n$th level.

In future experiments, to derive precise coupling parameters of Higgs boson to heavy quarks in these channels, one could take those high excited states contributions into account for greater dataset. Roughly, if all the high excited Fock states decay to the ground state $|(Q\bar{Q'})[1^1S_0]\rangle$ through electromagnetic or hadronic interactions, we can obtain the total decay width of Higgs boson to heavy quarkonia decay within the BT-potential model:
\begin{eqnarray}
\Gamma{(H^0\to |(b\bar{c})[1^1S_0]\rangle +\bar{c}b)} &=&25.10\;{\rm keV} \label{bc},\\
\Gamma{(H^0\to |(c\bar{c})[1^1S_0]\rangle +\bar{c}c)} &=&3.230\;{\rm keV} \label{cc},\\
\Gamma{(H^0\to |(b\bar{b})[1^1S_0]\rangle +\bar{b}b)} &=&2.359\;{\rm keV} \label{bb}.
\end{eqnarray}
Obviously, the decay width for $|(b\bar{c})[n]\rangle$ meson is larger than those of charmonium and bottomonium by about an order of magnitude.

At the HE-LHC, running at the center-of-mass energy of $\sqrt{s}=27$ TeV and producing a dataset corresponding to an integrated luminosity of $15$ $ab^{-1}$, the gluon-fusion cross-section of the Higgs boson production would be $151.6$ $pb$  \cite{lhc}. Then we can estimate the event numbers of $|(Q\bar{Q'})\rangle$ quarkonia production through Higgs boson decays, i.e, around $1.4~\times10^{7}$ of $(b\bar{c})$-meson,  $1.7~\times10^{6}$ of $(c\bar{c})$-meson and $1.3~\times10^{6}$ of $(b\bar{b})$-meson events can be obtained during the HE-LHC run. So, it is worth of the serious consideration to study $|(Q\bar{Q'})[n]\rangle$ quarkonia in these Higgs boson rare decays at the upgraded HE/HL-LHC and the newly purposed Higgs factories.

\subsection{Decay widths under five potential models}

\begin{table}
\caption{Decay widths (units: $keV$) for $|(b\bar{c})[n]\rangle$ quarkonium production channel $H^0\to |(b\bar{c})[n]\rangle+\bar{b}c$, where bound-state parameters under the five potential models ($n_f=4$) are adopted \cite{lx}.}
\begin{tabular}{|c|c|c|c|c|c|}
\hline
~~~~&~$BT$~&~$R$~&~$IO$~&~$CK$~&~$Cornell$~\\
\hline\hline
$[n]=[1S]$~&~14.16~&~6.461~&~18.59~&~5.111~&~6.298~\\
\hline
$[n]=[2S]$~&~2.580~&~2.894~&~3.616~&~2.059~&~2.944~\\
\hline
$[n]=[3S]$~&~1.853~&~2.168~&~1.18~&~1.501~&~2.244~\\
\hline
$[n]=[4S]$~&~1.694~&~1.794~&~1.131~&~1.228~&~1.886~\\
\hline
$[n]=[1P]$~&~2.207~&~0.966~&~1.375~&~0.610~&~0.639~\\
\hline
$[n]=[2P]$~&~0.779~&~0.959~&~0.732~&~0.618~&~0.679~\\
\hline
$[n]=[3P]$~&~0.922~&~0.867~&~0.427~&~0.507~&~0.639~\\
\hline
$[n]=[4P]$~&~0.907~&~0.863~&~0.309~&~0.530~&~0.652~\\
\hline
Sum~&~25.10 ~&~16.97~&~28.00~&~12.16~&~15.98\\
\hline\hline
\end{tabular}
\label{tabrpd}
\end{table}
\begin{table}
\caption{Decay widths (units: $eV$) for $|(c\bar{c})[n]\rangle$ quarkonium production channel $H^0\to |(c\bar{c})[n]\rangle+\bar{c}c$, where bound-state
parameters under the five potential models ($n_f=4$) are adopted \cite{lx}.}
\begin{tabular}{|c|c|c|c|c|c|}
\hline
~~~~&~$BT$~&~$R$~&~$IO$~&~$CK$~&~$Cornell$~\\
\hline\hline
$[n]=[1S]$~&~1211~&~515.3~&~309.6~&~410.9~&~503.4~\\
\hline
$[n]=[2S]$~&~569.6~&~381.2~&~238.7~&~273.1~&~372.4~\\
\hline
$[n]=[3S]$~&~349.8~&~331.6~&~211.8~&~231.6~&~325.6~\\
\hline
$[n]=[4S]$~&~298.6~&~307.3~&~198.4~&~212.4~&~303.9~\\
\hline
$[n]=[1P]$~&~286.8~&~117.7~&~49.70~&~70.62~&~79.34~\\
\hline
$[n]=[2P]$~&~141.6~&~145.9~&~64.01~&~85.55~&~104.0~\\
\hline
$[n]=[3P]$~&~183.2~&~160.9~&~72.22~&~92.59~&~118.3~\\
\hline
$[n]=[4P]$~&~188.7~&~170.0~&~77.63~&~96.74~&~127.4~\\
\hline
Sum~&~3230~&~2130~&~1222~&~1473~&~1934~\\
\hline\hline
\end{tabular}
\label{tabrpe}
\end{table}\begin{table}
\caption{Decay widths (units: $eV$) for $|(b\bar{b})[n]\rangle$ quarkonium  production channel $H^0\to |(b\bar{b})[n]\rangle+\bar{b}b$, where bound-state
parameters under the five potential models ($n_f=5$) are adopted \cite{lx}.}
\begin{tabular}{|c|c|c|c|c|c|}
\hline
~~~~&~$BT$~&~$R$~&~$IO$~&~$CK$~&~$Cornell$~\\
\hline\hline
$[n]=[1S]$~&~1037~&~4140~&~644.4~&~450.4~&~677.0~\\
\hline
$[n]=[2S]$~&~513.3~&~199.8~&~208.6~&~207.0~&~330.1~\\
\hline
$[n]=[3S]$~&~323.1~&~150.9~&~119.4~&~155.3~&~260.6~\\
\hline
$[n]=[4S]$~&~193.6~&~128.9~&~82.86~&~133.2~&~228.8~\\
\hline
$[n]=[1P]$~&~103.9~&~18.45~&~19.10~&~28.09~&~34.71~\\
\hline
$[n]=[2P]$~&~86.47~&~17.74~&~12.58~&~25.64~&~27.63~\\
\hline
$[n]=[3P]$~&~61.81~&~18.04~&~9.627~&~25.58~&~30.19~\\
\hline
$[n]=[4P]$~&~39.95~&~17.46~&~7.313~&~24.47~&~30.69~\\
\hline
Sum~&~2359~&~965.3~&~1012 ~&~1050~&~1620~\\
\hline\hline
\end{tabular}
\label{tabrpf}
\end{table}

For the leading-order calculation of the heavy $|(Q\bar{Q'})[n]\rangle$ quarkonium production and decay rates, their main uncertainty sources include the non-perturbative bound-state matrix elements, the  running coupling constant $\alpha_s$ and masses of heavy quarks and the Higgs boson. At present, values of the running coupling constant $\alpha_s$ and masses of the particles have been well restricted by experiments, so we shall not discuss them here. In the following, we will explore the uncertainty caused by the bound-state matrix elements, which are non-perturbative and model-dependent. We take the parameters derived under five potential models, i.e., the BT- potential~\cite{lx, pot2}, the QCD-motivated potential with one-loop correction given by Richardson (R-potential) \cite{Richardson:1978bt}, the QCD-motivated potential with two-loop correction given by Chen and Kuang (CK-potential) \cite{Chen:1992fq,Ikhdair:2003ry}, as well as by Igi and Ono (IO-potential) \cite{Igi:1986wg,Ikhdair:2003ry}, and Coulomb-plus-linear potential, also called the Cornell potential~\cite{Eichten:1978tg,Eichten:1995ch,Ikhdair:2003ry,lx}. The constituent quark masses and their corresponding radial wave functions at the origin and the first derivative of the radial wave functions at the origin for various $|(Q\bar{Q'})[n]\rangle$  Fock states can be found in tables I, II and III in our earlier manuscript \cite{lx}.

The decay widths for heavy $|(Q\bar{Q'})[n]\rangle$ mesons production in Higgs semi-exclusive decays under the five potential models are presented in Tables \ref{tabrpd}$\sim$\ref{tabrpf}.

\begin{itemize}
\item For the channel of $H^0\rightarrow |(b\bar{c})[n]\rangle+ c\bar{b}$,  the IO model gives the largest decay width among the five potential models, while the CK model gives the smallest values. Summing up the contributions of all Fock states and taking decay widths evaluated within the BT  model as the central value, we obtain its total decay width with uncertainties: $25.10^{+11.6\%}_{-51.6\%}$ keV.
\end{itemize}
\begin{itemize}
\item For charmonim production in $H^0$ semi-exclusive decays, the BT model gives the largest values and the IO model gives the smallest values. Summing up the contributions of all Fock states and taking decay widths evaluated under the BT model as the central value, we have $3.23^{+0\%}_{-62.2\%}$ keV for $H^0\rightarrow |(c\bar{c})[n]\rangle+ c\bar{c}$ channel.
\end{itemize}
\begin{itemize}
\item For bottomonium production in $H^0$ boson decays, the BT model gives the largest values and the R model gives the smallest ones. Summing up the contributions of all Fock states and taking decay widths evaluated within BT model as the central value, we obtain $2.36^{+0\%}_{-57.1\%}$ keV for $H^0\rightarrow |(b\bar{b})[n]\rangle+ b\bar{b}$ channel.
\end{itemize}
It is found that discrepancies caused by adopting different potential models can be as large as more than 50 percent.

\section{Conclusions}

In this work, we have made a comprehensive study on the high excited states of the $|(b\bar{c})[n]\rangle$ (or $|(c\bar{b})[n]\rangle$), $|(c\bar{c})[n]\rangle$ and $|(b\bar{b})[n]\rangle$ quarkonium production in Higgs boson semi-exclusive decays within the NRQCD factorization framework, i.e., $H^0\to |(b\bar{c})[n]\rangle +c\bar{b}$ (or $H^0\to |(c\bar{b})[n]\rangle +\bar{c}b$), $H^0\to |(c\bar{c})[n]\rangle +\bar{c}c$ and $H^0\to |(b\bar{b})[n]\rangle +\bar{b}b$ channels, where $[n]$ stands for $[n^1S_0]$, $[n^3S_1]$, $[n^1P_1]$, and $[n^3P_J]$, ($n=1, 2, 3, 4$; $J=0, 1, 2$). The``improved trace technology", which disposes the Dirac matrices at the amplitude level, is helpful for deriving compact analytical results especially for complicated processes with massive spinors. The total decay widths and differential distributions of $d\Gamma/ds_1$, $d\Gamma/ds_2$, $d\Gamma/dcos\theta_{12}$ and $d\Gamma/dcos\theta_{23}$ for above all Fock states are explored in detail. Further, for a sound estimation, we study the decay widths under five prevalent potential models and discuss the uncertainties.

According to the our study, numerical results show that the high excited Fock states of $|(Q\bar{Q'})[n]\rangle$ in addition to the ground $1S$ wave states can also provide sizable contributions to the heavy quarkonium production through Higgs boson decays, which implies that one could also consider exploring the coupling properties of Higgs boson to heavy quarks in these high excited states channels, especially for the charmonium and bottomonium. If almost all the high excited heavy quarkonium Fock states decay to the ground spin-singlet $1S$ wave state $|(Q\bar{Q'})[1^1S_0]\rangle$ through electromagnetic or hadronic interactions, we obtain the total decay width for $|(Q\bar{Q'})\rangle$ quarkonium production through $H^0$ semi-exclusive decays: $25.10^{+11.6\%}_{-51.6\%}$ keV for $|(b\bar{c})[n]\rangle$ meson, $3.23^{+0\%}_{-62.2\%}$ keV for $|(c\bar{c})[n]\rangle$ and $2.36^{+0\%}_{-57.1\%}$ keV for $|(b\bar{b})[n]\rangle$, where uncertainties are caused by varying the non-perturbative potential models. At the HE-LHC which runs at $\sqrt{s}=27$ TeV with an integrated luminosity of 15 $ab^{-1}$, the cross-section of the Higgs boson production in gluon fusion would be $151.6$ $pb$, hence we can obtain about $2.3~\times10^9$ Higgs boson events. Adopting total decay width of the Higgs boson $\Gamma_{H^0}=4.2$ MeV, sizable heavy quarkonium events can be produced through Higgs boson decays, i.e., about $1.4~\times10^{7}$ of $(b\bar{c})$ (or $(c\bar{b})$)-meson,  $1.7~\times10^{6}$ of $(c\bar{c})$-meson, and $1.3~\times10^{6}$ of $(b\bar{b})$-meson events can be obtained.

\hspace{2cm}

{\bf Acknowledgements}: We thank Professor Xing-Gang Wu for helpful discussions. This work was supported in part by the Shandong Provincial Natural Science Foundation, China (ZR2019QA012) and the Fundamental Research Funds of Shandong University (2019GN038).

\end{document}